\documentclass[useAMS,usenatbib]{mn2e}

\usepackage{amssymb}
\usepackage{amsmath}
\usepackage{times}
\usepackage{graphicx}

\newcommand{\lya}{Ly$\alpha$}

\newcommand{\kms}{\mbox{km s$^{-1}$}}

\newcommand{\apjs}{ApJS}
\newcommand{\apj}{ApJ}
\newcommand{\aj}{AJ}
\newcommand{\araa}{ARA\&A}
\newcommand{\apjl}{ApJL}
\newcommand{\nat}{Nature}
\newcommand{\pasp}{PASP}
\newcommand{\aap}{A\&A}
\newcommand{\mnras}{MNRAS}
\newcommand{\procspie}{Proc. SPIE}

\def\ltsima{$\; \buildrel < \over \sim \;$}
\def\simlt{\lower.5ex\hbox{\ltsima}}
\def\gtsima{$\; \buildrel > \over \sim \;$}
\def\simgt{\lower.5ex\hbox{\gtsima}}

\title[Rest-Frame UV Spectrum of 'The Cosmic Horseshoe']
{The Ultraviolet Spectrum of the Gravitationally Lensed Galaxy 
`The Cosmic Horseshoe': A Close-up of a Star-forming Galaxy 
at $z \sim 2$.}

\author[Quider et al.]{Anna M. Quider$^{1}$\thanks{Email: aquider@ast.cam.ac.uk}, Max Pettini$^1$, Alice E. Shapley$^2$, and Charles C. Steidel$^3$\\ 
$^1$ Institute of Astronomy, Madingley Rd, Cambridge, CB3 0HA, UK\\ 
$^2$ Department of Physics and Astronomy, University of California, Los Angeles, 
CA 90095-1547, USA\\ 
$^3$ California Institute of Technology, Mail Stop 105-24, Pasadena, CA 91125, USA}

\begin{document}

\date{Accepted ... Received ... in original form ...}

\pagerange{\pageref{firstpage}--\pageref{lastpage}} \pubyear{2009}

\maketitle

\label{firstpage}

\begin{abstract}
Taking advantage of strong gravitational lensing, we have
recorded the rest-frame ultraviolet spectrum 
of the $z = 2.38115$ galaxy `The Cosmic
Horseshoe' (J1148+1930) at higher resolution and 
signal-to-noise ratio than is currently feasible for 
unlensed galaxies at $z = 2 -3$.
With a star-formation rate of $\textrm{SFR}\sim 100 M_\odot$~yr$^{-1}$,
dynamical mass $M_{\rm vir} \simeq 1 \times 10^{10} M_\odot$,
half-solar metallicity, and moderate reddening
$E(B-V) = 0.15$, the Cosmic Horseshoe is a good
example of the population of galaxies responsible for
most of the star-formation activity at these redshifts.

From the analysis of stellar spectral features we conclude
that a continuous mode of star formation with a 
Salpeter slope for stars in the mass range 5--$100 M_\odot$
gives a good representation of the UV spectrum, ruling out
significant departures from a `standard' IMF.
Generally, we find good agreement between the values
of metallicity deduced from stellar and nebular tracers.
Interstellar absorption is
present over a velocity range 
$\Delta v \simeq 1000$\,km~s$^{-1}$, 
from $-800$ to +250\,km~s$^{-1}$
relative to the stars and their H\,{\sc ii} regions, but
we still lack a model relating the kinematic structure of the
gas to its location within the galaxy. 
There is evidence to suggest that the outflowing interstellar gas
is patchy, 
covering only $\sim 60$\% of the UV stellar continuum. 

The \lya\ line shares many of the characteristics of the 
so-called \lya\ emitters; its double-peaked profile can be 
reproduced by models of \lya\
photons resonantly scattered 
by an expanding shell of gas and dust, with
$\sim 10$--15\% of the photons escaping the galaxy.
Many of the physical properties of the Cosmic Horseshoe are 
similar to those of the only other galaxy at
$z = 2$--3 studied in comparable detail up to now: MS\,1512-cB58.
The fact that these two galaxies have drastically different \lya\
lines may be due simply to orientation effects, or differences in
the covering factor of outflowing gas, and cautions
against classifying high-$z$ galaxies only on the basis of  
spectral features, such as \lya, whose appearance
can be affected by a variety of different parameters. 

\end{abstract}

\begin{keywords}
cosmology: observations --- galaxies: evolution --- galaxies: starburst --- galaxies: individual (Cosmic Horseshoe)
\end{keywords}

\section{Introduction}
\label{sec:introduction}

In the thirteen years that have elapsed since the term ``Lyman break galaxy''
(or LBG) was coined, 
samples of galaxies at $z = 2$--4 have increased a thousand-fold.
Large surveys, selecting galaxies not only
via the break at 912\,\AA\
(due mostly to intergalactic H\,{\sc i} absorption), 
but also via \lya\ emission, a variety of colour
criteria, and sub-mm emission,  have given us a broad
view of the galaxy population during the epoch when star-formation
activity was at its peak in the history of our Universe 
(see Pettini et al. 2008 for a brief review). 

However, typical galaxies at $z \simeq 3$ are faint, 
with fiducial ${\cal R}$-band magnitude 
$m_{\cal R}^{\ast} = 24.4$ (Steidel et al. 1999; Reddy et al. 2008),
so that, even with the most efficient optical and infrared (IR)
telescopes and spectrographs currently
available,  it is only possible to record their spectra 
with limited resolution and signal-to-noise ratio.
Until the next generation of 30+\,m optical/IR
telescopes comes into operation, the only strategy at
our disposal for studying \textit{in detail} the internal
properties of normal galaxies at these redshifts is to take
advantage of fortuitous alignments with foreground
mass concentrations (massive galaxies or galaxy clusters) 
which can provide an order-of-magnitude
boost of the flux reaching the Earth through gravitational
lensing. 

For several years, the archetypal object for this kind of
study has been the Lyman break galaxy MS\,1512-cB58
(or cB58 for short; Yee et al. 1996).
The magnification by a factor of $\sim 30$ (Seitz et al. 1998)
provided by an intervening galaxy cluster at $z = 0.37$ 
brings this $L^{\ast}$, $z = 2.7276$, galaxy within reach of 
high resolution spectroscopy from the ground and has made
it the target of extensive observations at wavelengths
from the visible to the mm range 
(Pettini et al. 2000, 2002;  Savaglio, Panagia, \& Padovani 2002;
Teplitz et al. 2000; Siana et al. 2008; 
Baker et al. 2001, 2004; Sawicki 2001).
These studies provided unprecedented clear views of the 
interstellar medium, young stars, star-formation
history, dust, metallicity, kinematics, and many other
physical properties of a `typical' LBG at $z \sim 3$.

But how typical is cB58?
We know from the work of Shapley et al. (2003)
that its ultraviolet (UV) continuum is more reddened, 
and its interstellar absorption lines are stronger,
than in an `average' LBG.
Furthermore, its young age of only 10--20\,Myr
(Ellingson et al. 1996; Siana et al. 2008)  
is at the lower end of the wide range of values 
deduced by fitting the spectral energy distributions
of galaxies at $z = 2$--3 (Papovich, Dickinson, \& Ferguson
2001; Shapley et al. 2001; Erb et al. 2006b),
although it is likely that an older stellar population 
is present too.
Clearly, it is important to study in detail
several other similarly bright sources
and establish the \emph{range} of properties
of galaxies at a given cosmic epoch, rather than  
naively assuming that any particular galaxy is
representative of a whole population. Fortunately, 
the large area of sky surveyed by the 
\textit{Sloan Digital Sky Survey (SDSS)} has led to
the recent identification of many other examples of 
highly magnified galaxies (see, for instance, Belokurov
et al. 2009; Kubo et al. 2009, and references therein).
These and other discoveries have spurred a 
number of follow-up studies
(e.g. Smail et al. 2007; Stark et al. 2008; Finkelstein et al. 2009;
Hainline et al. 2009; Yuan \& Kewley 2009).

With these new samples of objects it will be possible to
make progress on a variety of outstanding questions which cannot
be satisfactorily addressed with lower resolution data.
In particular:

(i) To what extent do different metallicity indicators,
based on stellar photospheric lines, wind lines from the 
most massive stars, interstellar absorption lines,
and emission lines from H\,{\sc ii} regions, give 
consistent answers? This is an important issue,
not only to explore the degree of chemical homogeneity
of galaxies undergoing rapid star-formation, but also
to clarify the sources of the systematic offsets 
between different H\,{\sc ii} region
metallicity calibrators, and possibly bring them into
better internal agreement (see, for example,  the discussions
by Pettini 2006, and Kewley \& Ellison 2008).

(ii) What do the profiles of the interstellar absorption lines
tell us about large-scale outflows in actively star-forming
galaxies and about the inflow of gas fueling star formation?
Can we see any evidence of the leakage of hydrogen
ionizing photons from the sites of star formation
into the intergalactic medium (IGM) which has proved
so difficult to detect directly 
(e.g. Shapley et al. 2006; Iwata et al. 2009)
and yet seems to be required by a number of indirect
lines of evidence (e.g. Faucher-Gigu{\`e}re et al. 2008)?

(iii) Can we place limits on possible variations of the 
stellar initial mass function (IMF) at $z = 2$--3? Claims
to this effect have certainly been put forward
(e.g. Wilkins et al. 2008 and references therein),
but the evidence is still controversial
(e.g. Reddy \& Steidel 2009).

(iv) What are the factors affecting the wide variety
of spectral morphologies of the \lya\ line, from
strong narrow emission to damped absorption
(e.g. Mas-Hesse et al. 2003; Verhamme, Schaerer, \& Maselli 2006),
and what can we learn about the relationship
of the so called `\lya\ emitters' to the more 
general LBG population?

In this paper we address these and other related questions
with new observations of the rest-frame UV 
spectrum of a recently identified highly
magnified LBG, named `The Cosmic Horseshoe' by its
discoverers (Belokurov et al. 2007). The paper is organised
as follows.
In Section 2, we summarize the known properties
of the Cosmic Horseshoe and in Section 3 
we provide details
of our observations and data reduction.
In Section 4, we consider the information provided
by the stellar UV spectrum of this galaxy, concerning in
particular its systemic redshift, massive star IMF,
and metallicity of the OB stars.
Section 5 deals with the resolved profiles of the
interstellar absorption lines, while Sections~6 and 7
focus on the UV nebular emission lines covered
by our spectrum, \lya\ and C\,{\sc iii}]\,$\lambda 1909$
respectively.
In Section 8, we bring these different strands
together in the light of the questions
raised above. Finally, we summarize our main
conclusions in Section 9.
Throughout the paper, we adopt a cosmology
with $\Omega_{\rm M} = 0.3$, $\Omega_{\Lambda} = 0.7$,
and $H_0 = 70$\,km~s$^{-1}$~Mpc$^{-1}$.

\section{The Cosmic Horseshoe}

\begin{figure}
\vspace*{0.25cm}
\centerline{
\includegraphics[width=0.85\columnwidth,clip,angle=0]{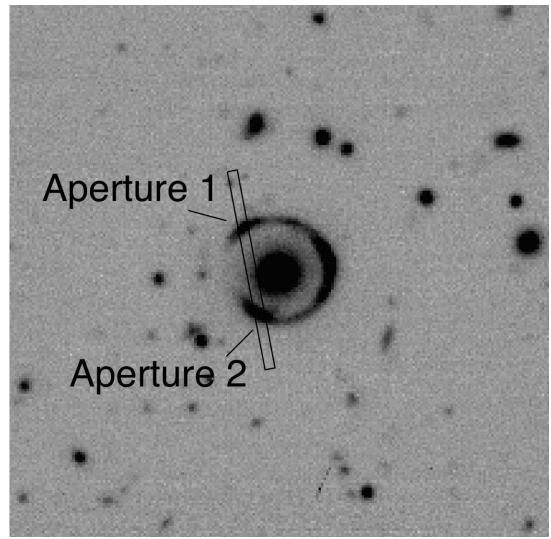}}
\caption{$R$-band image of the Cosmic Horseshoe 
obtained with FORS2 on the Very Large Telescope of the European
Southern Observatory (courtesy of L.~J. King).
North is up and East is to the left. 
Superposed on the image is the $1.0 \times 20$\,arcsec entrance
slit of ESI used for the observations reported here.
}
\label{fig:horseimage}
\end{figure}

The Cosmic Horseshoe, or J1148+1930,
is a nearly complete blue Einstein ring
of 10\,arcsec diameter encircling a red 
galaxy (Figure~\ref{fig:horseimage}; see also  
http://www.ast.cam.ac.uk/research/cassowary/).
Low resolution
spectroscopy by Belokurov et al. (2007) revealed the 
lens to be a massive luminous red galaxy 
($M \sim 6 \times 10^{12} M_{\odot}$) at $z = 0.444$
and the ring to be the gravitationally lensed image 
of a star-forming galaxy at $z = 2.379$.
From the extent of the ring, Belokurov et al.
estimated an approximate magnification factor of $\sim 35$,
while the detailed lensing models constructed
by Dye et al. (2008) give overall magnification factors 
of $24 \pm 2$.
Adopting the latter value, and 
interpolating between the 
total integrated magnitudes of the
ring $g = 20.1$, $i = 19.7$ reported
by Belokurov et al. (2007), we deduce a
luminosity $L \simeq 2.4 L^{\ast}$
for the source, relative to 
$M_{\rm AB}^{\ast}(1700\,{\rm \AA}) = -21.0$ from 
the luminosity function of UV-selected $z \simeq 2$ galaxies
by Reddy et al. (2008). 
The image reconstruction in the source plane with the 
best fitting lensing model 
by Dye et al. (2008) shows that the major contributor to this
luminosity is a compact source with half-light radius
$r_{1/2} \sim 0.3$\,arcsec, or $\sim 2.5$\,kpc 
at $z = 2.379$; there is also evidence for a second, fainter, source 
$\sim 0.7$\,arcsec to the North.

We briefly review the information 
currently available on the lensed galaxy. 
Its rest-frame UV spectrum 
is essentially flat in $F_{\nu}$, reflecting a 
recent star-formation episode and modest 
reddening [$E(B-V) \simeq 0.1$--0.2; Belokurov et al. 2007],
but a more extensive characterization of its 
stellar population(s) awaits longer 
wavelength photometry to build-up the spectral
energy distribution  from the UV to the near-IR
(e.g. Shapley et al. 2005; Erb et al. 2006b).
Recently, Hainline et al. (2009) reported observations
of several rest-frame optical emission lines formed in the
H\,{\sc ii} regions of the Cosmic Horseshoe, from
which they were able to measure a number of
parameters of interest. 
The H$\alpha$ luminosity implies
a star formation rate of $113\,M_{\odot}$~yr$^{-1}$,
after correcting for reddening and a lensing magnification
factor of 24.
The emission line widths indicate a velocity dispersion
$\sigma \simeq 65$\,km~s$^{-1}$ and a corresponding
virial mass $M_{\rm vir} \simeq 1.0 \times 10^{10} M_{\odot}$.
The oxygen abundance deduced from the ratios of the strongest
emission lines is in the range 
(O/H)$_{\rm H\,{\textsc{ii}}} \simeq 0.5$--1.5\,(O/H)$_{\odot}$.
These parameters are typical of UV-bright galaxies at
$z \sim 2$ (Erb et al. 2006a,b,c), although the star formation
rate is at the upper end of the distribution of values
found by Erb et al. (2006c).
It is of interest now to compare such estimates with 
those provided by the rest-frame UV spectrum.

\section{Observations and Data Reduction}
\label{sec:observations}

We observed the Cosmic Horseshoe with 
the Echellette Spectrograph and Imager (ESI; Sheinis et al. 2000) 
on the Keck\,{\sc ii} telescope.  
The high efficiency, moderately high resolution
($R = 4000$ with a 1\,arcsec slit, sampled
with 11.4\,km~s$^{-1}$ pixels),
and wide spectral coverage of 
ESI---from 4000 to 10\,000\,\AA, or 1184--2959\,\AA\ in the
rest-frame of the Cosmic Horseshoe---make it highly
suitable for an in-depth study of the UV spectrum of
this galaxy.
The data were collected over two nights (2008 March 6 and 7 UT) in 
mostly sub-arcsecond seeing; the total exposure time was 36100\,s,
made up of a number of 1800\,s or 2000\,s-long individual integrations. 
The ESI slit was oriented at sky P.A. = 10.8 degrees East of North,
and positioned so as to encompass the two brightest knots
in the Einstein ring, as shown in Figure~\ref{fig:horseimage}.
The northern knot, labelled `Aperture 1' in the Figure, is
knot `A' in the labelling of the features by Belokurov et al. (2007),
while `Aperture 2' corresponds to their knot `D'.  
Both knots are due to the same object in the source plane,
although `Aperture 1' may include a small contribution from
the fainter component $\sim 0.7$\,arcsec to the North of the main
source of the ring (Dye et al. 2008). The location of the ESI slit is the same
as that of the NIRSPEC observations by Hainline et al. (2009).

The data were processed with standard IRAF tasks. 
Each individual two-dimensional (2-D) ESI spectrum
was geometrically corrected (to correct for the spectral
curvature of each echellette order), 
bias subtracted, flat-fielded, and corrected for cosmetic defects and 
cosmic-ray affected pixels.
In the next step, all of the images thus processed were co-added
into a total 2-D frame which was then background subtracted.
The background subtraction process included careful modelling of
the distribution along the slit of light from the lens, which is clearly
visible above the sky at the longer wavelengths of the 
ESI range. Finally, two 1-D spectra were extracted
(using optimal extraction algorithms), one
for each `Aperture'.  The extraction process included
merging the 10 echellette orders of ESI
into one continuous spectrum, with weighted addition of
portions of adjacent orders which overlap in wavelength.

We compared carefully the 1-D spectra of Aperture 1 and
Aperture 2 to look for differences which one may expect
to be revealed by the gravitational magnification of any
spatial structure in the source, but found none. Neither
dividing one spectrum by the other nor subtracting one
from the other revealed features more significant than
the noise.
As discussed in Section~\ref{sec:lya}, 
this is also the case for the \lya\ emission line which 
(a) has the highest signal-to-noise ratio (S/N), and
(b) is most sensitive to geometrical effects, 
being resonantly scattered.
The only difference between the two apertures is in the
total flux: emission line and continuum are uniformly
higher by a factor of 1.19 in Aperture 2 than in Aperture 1.
Thus, in order to improve the S/N, the two spectra were
averaged and the resulting spectrum mapped
onto 0.5\,\AA\ bins. 

This final spectrum has a resolution (full width at half maximum)
FWHM\,=\,75\,km~s$^{-1}$, sampled with three wavelength
bins at 6000\,\AA, as determined from the widths of narrow
emission lines from the Cu-Ar and Hg-Ne-Xe 
hollow-cathode lamps used
for wavelength calibration, whose spectrum was processed
in the same way as that of the Horseshoe.
From the rms deviation of the data from the mean,
we measured an average S/N\,$\simeq 8$ per 0.5\,\AA\ 
wavelength bin
between $\sim 4000$ and $\sim 7500$\,\AA\
($\sim 1200$ and $\sim 2200$\,\AA\ in the rest frame),
with a factor of $\sim 50$\% variation in S/N across each echellette
order. At the wavelengths of the \lya\ emission line the S/N
rises to a maximum S/N\,=\,44 per 0.5\,\AA\ bin.

We attempted to put our spectrum on an absolute flux scale
by reference to those of three flux-standard stars 
recorded during the
two nights of observation, but found a scatter of 
$\pm 20$\% in the flux calibration among 
the three stars. 
Adopting the mean value of the three flux scales,
we deduced a mean 
$\langle f_\nu \rangle = 8.3 \times 10^{-29}$\,erg~s$^{-1}$~cm$^{-2}$~Hz$^{-1}$
between 4200--5485\,\AA.
This wavelength interval corresponds to the 
FWHM of the transmission curve of the Gunn $g$ filter
through which Belokurov et al. (2007) measured $g = 20.1$
(corresponding to 
$\langle f_\nu \rangle = 3.3 \times 10^{-28}$\,erg~s$^{-1}$~cm$^{-2}$~Hz$^{-1}$)
for the entire Einstein ring.
Thus, if the flux calibration is correct, the 
ESI slit captured a fraction $0.25 \pm 0.05$ of the
light of the Cosmic Horseshoe, corresponding
to an effective magnitude within the two 
Apertures of $g = 21.6 \pm 0.2$.

\section{The Stellar Spectrum}
\label{sec:stellar_spectrum}
The rest-frame UV spectra of star-forming galaxies are complex
blends of interstellar absorption lines, nebular emission lines,
and absorption/emission lines formed in the atmospheres of 
OB stars. 
This wealth of spectral features, properly interpreted,
is a rich source of information on the physical properties
of the gas and stars in these galaxies. 
In this section, we consider photospheric absorption lines
which have been shown to be useful abundance diagnostics,
and P-Cygni profiles formed in the expanding atmospheres
of the most luminous early-type stars whose strength is
also sensitive to the upper end of the initial mass function.

\subsection{The systemic redshift of the Cosmic Horseshoe}
\label{sec:z_sys}

The first step in the analysis is the determination of the
systemic redshift of the galaxy.
While most photospheric features are unresolved blends
of multiple lines, a few exceptions allow the redshift
of the hot stars to be determined. Within the wavelength
range covered by our data, we identified the following
unblended photospheric absorption lines:
S\,{\sc v}\,$\lambda 1501.763$, 
N\,{\sc iv}\,$\lambda 1718.551$, 
and C\,{\sc iii}\,$\lambda 2297.579$ 
[$\lambda_{\rm air} = 2296.871$---all three vacuum wavelengths
from the National Institute of Standards and Technology (NIST)
database available at http://physics.nist.gov/PhysRefData/ASD/].

The redshifts of these three lines are in good mutual agreement
(see Table~\ref{table:systemic_z}), and define an average
$z_{\rm stars} = 2.38115 \pm 0.00006$. The $1 \sigma$
error corresponds to an uncertainty of $\pm 5$\,km~s$^{-1}$, or
1/5 of a wavelength bin in our final spectrum.
Also included in Table~\ref{table:systemic_z} are values
of $z_{\rm H\,\textsc{ii}}$ from the 
C\,{\sc iii}]\,$\lambda\lambda 1906.683, 1908.734$ 
doublet lines which
are resolved in our ESI spectrum (see Section~\ref{sec:CIII}), 
and from H$\alpha$ 
measured from the 
NIRSPEC spectrum of Hainline et al. (2009).
The mean $z_{\rm C\,\textsc{iii}]} = 2.38115 \pm 0.00012$ is
in excellent agreement with $z_{\rm stars}$; 
$z_{\rm H\alpha}$ differs by only 7\,km~s$^{-1}$ which 
may include a small systematic offset between the
wavelength calibrations of the ESI and NIRSPEC  spectra.
Thus, we conclude that the systemic redshift of the Cosmic Horseshoe
is $z_{\rm sys} = 2.38115$ and we adopt this value throughout
the paper.

\begin{table}
\centering
{\hspace*{1.35cm}\begin{minipage}[c]{1.0\textwidth}
 \caption{\textsc{Systemic Redshift}}
   \begin{tabular}{llllll} 
\hline 
\hline
Ion & $\lambda_{\rm lab}^{\rm a}$ (\AA) & $z_{\rm abs}$ & Origin\\ 
\hline 
S~{\sc v}   & 1501.763  & 2.38107   &   Stars \\
N~{\sc iv} & 1718.551  & 2.38119   &  Stars\\
C~{\sc iii}] & 1906.683 & 2.38103   &  H\,{\sc ii} Regions\\
C~{\sc iii}] & 1908.734 & 2.38127   &  H\,{\sc ii} Regions\\
C~{\sc iii}  & 2297.579  & 2.38118  &  Stars\\
H$\alpha^{\rm b}$ & 6564.614 & 2.38123 & H\,{\sc ii} Regions\\
\hline
     \label{table:systemic_z}
 \end{tabular}

 $^{\rm a}$ Vacuum wavelengths\\
 $^{\rm b}$ As reported by Hainline et al. (2009)
 \end{minipage}
 }
 \end{table}

\subsection{Photospheric lines}
\label{section:R04Indices}

In this and the following subsection, we analyse photospheric and wind lines
in the spectrum of the Cosmic Horseshoe by comparing our data with 
model spectra computed with the population synthesis code
\textit{Starburst99} which couples libraries of either empirical
(Leitherer et al. 1999, 2001) or theoretical (Rix et al. 2004) ultraviolet
OB stellar spectra with stellar evolutionary tracks. The input set of
parameters
to \textit{Starburst99} can be adjusted to simulate a variety
of star formation histories, ages, IMF parameters, and metallicities
so as to determine the combination that best fits the observations
under scrutiny.

\begin{figure*}
\centerline{
{\hspace*{-0.15cm}\includegraphics[width=1.55\columnwidth,clip,angle=270]{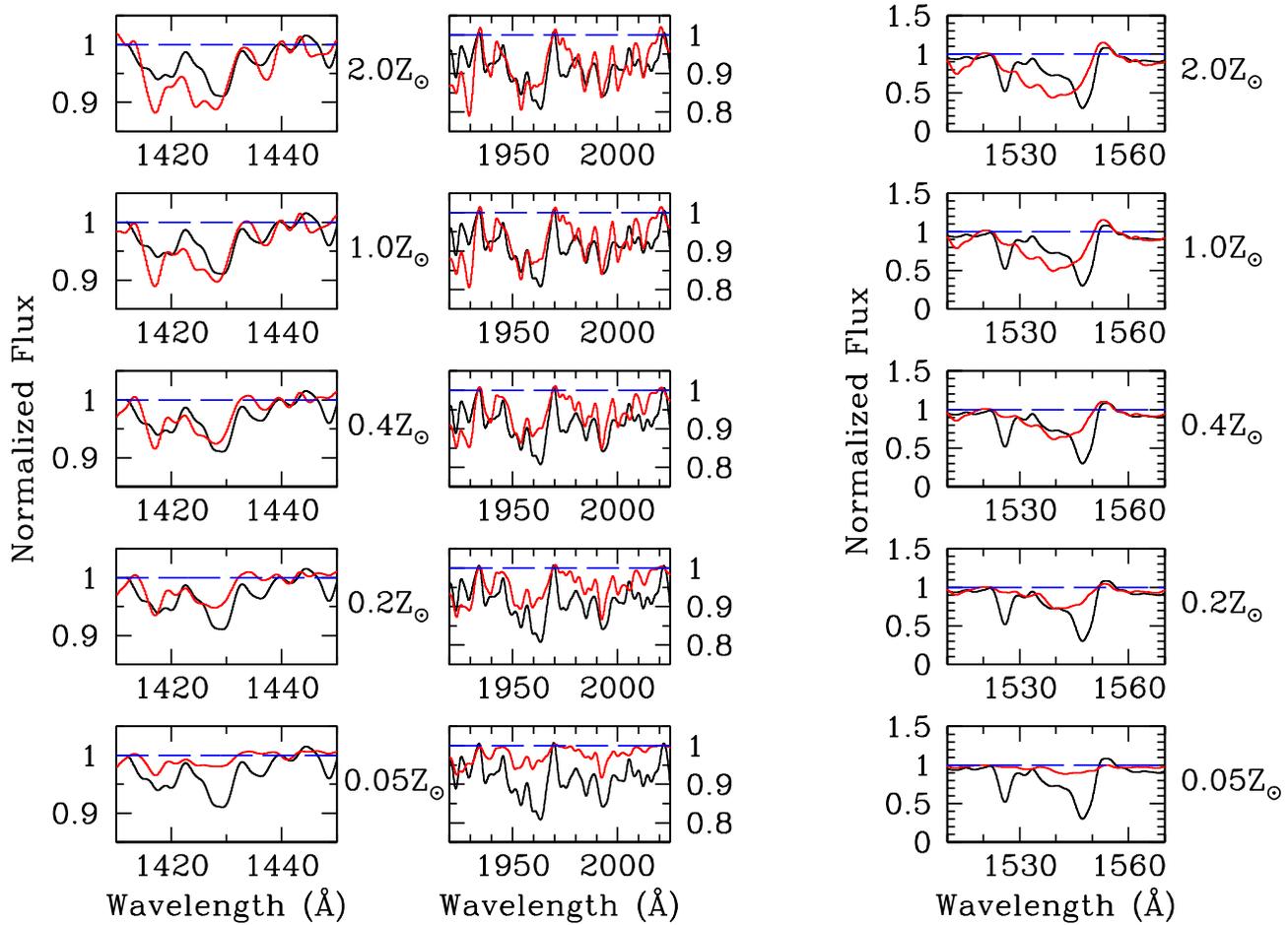}}
}
\caption{\textit{Left-hand pair of panels:} 
Portions of the UV spectrum of the Cosmic Horseshoe 
(black) in the regions encompassing the 1425 and 1978 
indices compared with model spectra (red) from the work by
Rix et al. (2004) for five different metallicities, as indicated.
The ESI spectrum has been smoothed to the 2.5\,\AA\ resolution
of the model spectra. See text for further details of the models.
\textit{Right-hand panel:} Comparison of the Rix et al. models
with the smoothed and normalized
spectrum of the Cosmic Horseshoe in the region of the
C\,{\sc iv}\,$\lambda \lambda 1548.204, 1550.781$ doublet. This region also includes
C\,{\sc iv}   and Si\,{\sc ii}\,$\lambda 1526.7070$
\textit{interstellar} absorption which is unaccounted for in the
Rix et al. models. Note the difference in the $y$-axis scales
between the C\,{\sc iv} panels and those for the weaker
photospheric 1425 and 1978 features to the left.
 }
\label{fig:RixIndices}
\end{figure*}

In the absence of more detailed information on the previous
history of star formation in the Cosmic Horseshoe, we shall
adopt the simplest case of continuous star formation---with 
a Salpeter slope of the upper end of the IMF---which has 
been proceeding at a
steady rate (as opposed to a declining rate, or a series of
bursts) for 100\,Myr (this being a safe lower limit beyond
which the spectrum no longer changes with time because the stars
that contribute to the UV light are born and die at the same rate).
While this scenario is undoubtedly an over-simplification, it may
not be too different from reality when considering the spectrum
of a whole galaxy, rather than individual 
regions of star formation. An age of more than 100\,Myr
 is typical of most $z = 2$ UV-bright galaxies in the
`BX' sample of Erb et al. (2006b) and seems plausible given the
near-solar metallicity of the H\,{\sc ii} regions of the Horseshoe
(Hainline et al. 2009). 

Leitherer et al. (2001) and Rix et al. (2004) pointed out the existence
of some blends of UV photospheric lines whose strengths, 
under the above assumptions, depend on metallicity
(see also Halliday et al. 2008). 
The ``1425'' index of Rix et al. (2004) measures
the equivalent width of a blend of 
Si\,{\sc iii}\,$\lambda 1417$, C\,{\sc iii}\,$\lambda 1427$, and 
Fe\,{\sc v}\,$\lambda 1430$ absorption lines spanning the
wavelength interval 1415--1435\,\AA;
at longer wavelengths, the ``1978'' index measures the 
strengths of several Fe\,{\sc iii} absorption lines 
from B stars between 1935 and 2020\,\AA.
Both indices stabilise after $\sim 50$\,Myr from the onset
of star formation, and increase monotonically with
metallicity.
In the left-hand portion of Figure~\ref{fig:RixIndices},
we compare the ESI spectrum
of the Cosmic Horseshoe with the model spectra of Rix et al. (2004)
for five values of metallicity, from 1/20 of solar to twice solar,
in the regions of the 1425 and 1978 indices.  For this comparison,
we smoothed our spectrum to the 2.5\,\AA\ resolution of
the Rix et al. models, and normalized it using the psuedo-continuum
windows suggested by those authors. Thus, the observed and model 
spectra should be directly comparable. 

Figure~\ref{fig:RixIndices} shows a generally good agreement
in the detailed features of the synthetic and real spectra in the two
wavelength regions considered here, attesting to the sophistication
reached by current hot-star atmosphere and stellar
population synthesis models. 
Referring to the 1425\,\AA\ region (left-most panel in Figure~\ref{fig:RixIndices}),
it can be seen that the strength of the photospheric lines in this
region of the Cosmic Horseshoe spectrum is intermediate
between those computed with metallicities 
$Z = 0.4 Z_{\odot}$  and $Z = 1.0 Z_{\odot}$
respectively. Using the relationship between metallicity and
the 1425 index proposed by Rix et al. (2004):
\begin{equation}
\log(Z/Z_{\odot}) = A \times {\rm EW(1425)} + B
\label{eq:1425index}
\end{equation}
with $A = 1.14$ and $B = -1.75$ appropriate to the value
${\rm EW}(1425) = 0.95$\,\AA\ we measure here, we derive
$Z_{\rm OB stars} = 0.5 Z_{\odot}$.

Qualitatively, the observed spectrum seems to be intermediate
between the 0.4 solar and solar metallicity cases in the 1978 region
too (see middle panel of Figure~\ref{fig:RixIndices}). 
However, there appears to be  
excess absorption between 1960 and 1980\,\AA\
over that expected from the models for \emph{any} metallicity;
the presence of these additional features does not allow us
to deduce a value of $Z_{\rm B stars}$ from EW(1978)
using the Rix et al. (2004) calibration of this index.
We considered the possibilities that the excess absorption may
be noise, or an intervening absorption system unrelated to the
Cosmic Horseshoe, but could find no evidence to support either
interpretation. 
It also seems unlikely that the excess is
due to a different mix of stellar spectral types or
chemical elements, since all of the photospheric absorption
in this region is thought to be due to Fe\,{\sc iii} lines,
and the rest of the spectral features between 1920 and 2025\,\AA\
match the data well. 
No such discrepancies were found by Rix et al. (2004)
in the two galaxies they considered 
(MS\,1512-cB58 and Q1307-BM1163, the latter
at $z = 1.411$), nor by Halliday et al. (2008)
in their composite spectrum of $z \sim 2$ galaxies.
However, the cases where these photospheric indices have
been measured are still very few,
and it is important to continue to look critically at the match
between model and real spectra as more data of 
suitable quality become available.

\begin{figure*}
\vspace{-1.75cm}
\centerline{
\includegraphics[width=1.65\columnwidth,clip,angle=270]{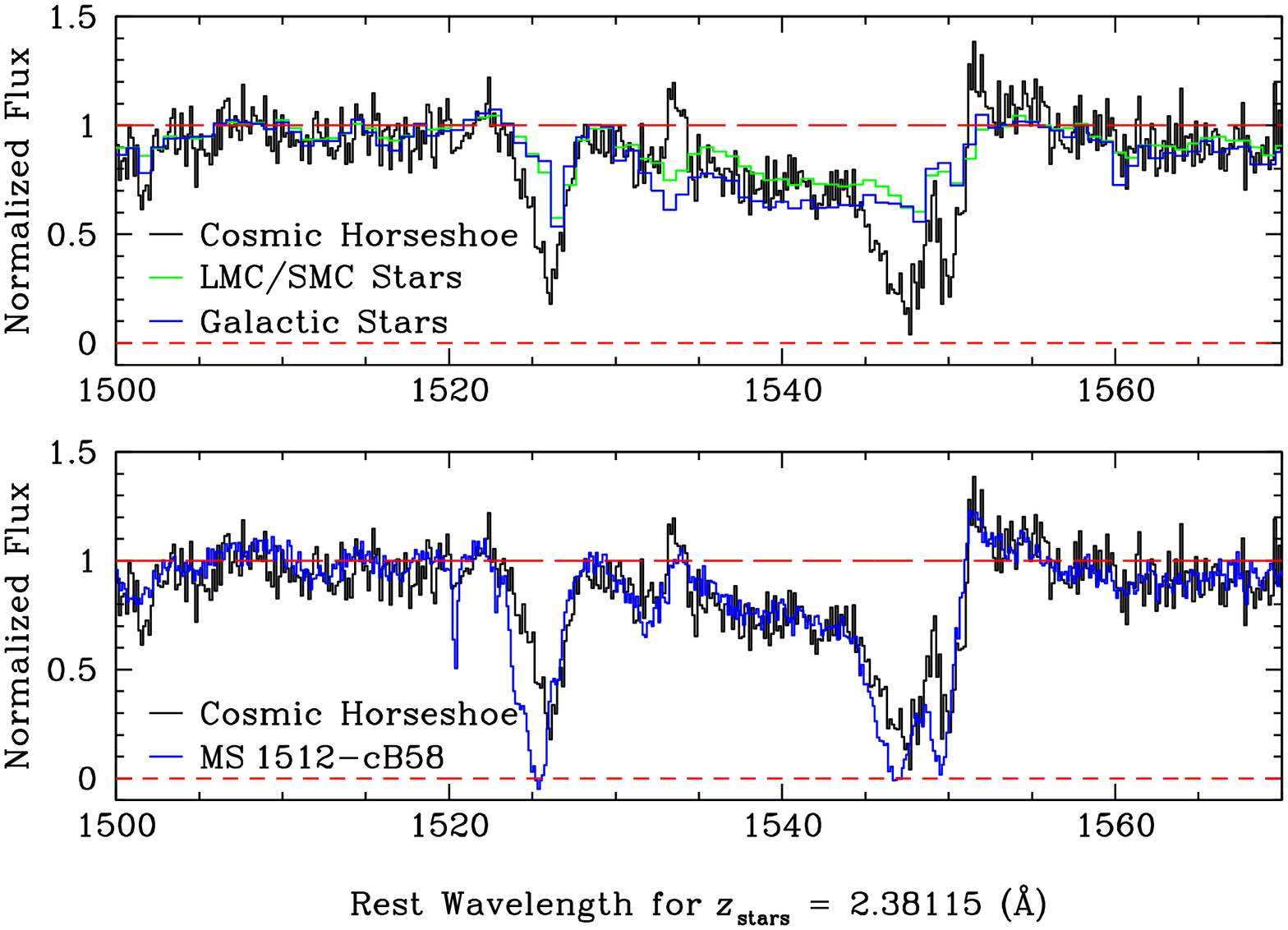}}
\vspace{-0.2cm}
\caption{\textit{Upper panel:}
Comparison between the ESI spectrum of the Cosmic Horseshoe
in the region encompassing the C\,{\sc iv} line and model spectra
computed with \textit{Starburst99} and empirical libraries
of Galactic and Magellanic Clouds stars, as indicated.
The model spectra were generated assuming a 100\,Myr
old continuous star formation episode with a Salpeter IMF.
\textit{Lower panel:} The stellar spectra of the Cosmic Horseshoe
and of MS\,1512-cB58 are remarkably similar in the wavelength
region shown. The ESI spectrum of cB58 is reproduced from Pettini
et al. (2002), and has been reduced to its rest wavelengths
at $z_{\rm stars} = 2.7276$.
}
\label{fig:CIV}
\end{figure*}

\subsection{Wind lines}
\label{section:stellarwind}

The model spectra by Rix et al. (2004) show that the spectral
features most sensitive to metallicity are the P-Cygni lines
formed in the expanding winds of the most luminous OB stars
whose mass-loss rates are thought to be lower at lower
metallicities (e.g. Kudritzki \& Puls 2000). 
Among the P-Cygni lines covered by our spectrum, 
the C\,{\sc iv}\,$\lambda \lambda 1548.204, 1550.781$ (unresolved)
doublet is the strongest.
Referring to the rightmost panel
in Figure\,\ref{fig:RixIndices}, we 
find that its strength in the Cosmic Horseshoe
is intermediate between those of the models
with metallicities $Z = 0.2 Z_{\odot}$ and $Z = 0.4 Z_{\odot}$.
However, as emphasised by Crowther et al. (2006),
the interpretation of this feature is complicated
by its blending with interstellar C\,{\sc iv}
(and to a lesser extent Si\,{\sc ii}\,$\lambda 1526.7070$) absorption,
at least at the coarse resolution of most spectra
of distant galaxies. Similar problems affect the
generally weaker Si\,{\sc iv}\,$\lambda\lambda 1393.7602, 1402.7729$
and N\,{\sc v}\,$\lambda 1238.821, 1242.804$ wind lines.

At the high resolution of our ESI spectrum,
interstellar and stellar components are easily separated,
as can be appreciated from Figure~\ref{fig:CIV}. 
The top panel compares the observed spectrum 
of the Cosmic Horseshoe in the C\,{\sc iv} region
with those generated by \textit{Starburst99}
now using libraries of \emph{empirical} stellar spectra,
assembled from UV observations of stars in either the Galaxy
or  the Magellanic Clouds. Both models are for 100\,Myr old
continuous star formation with Salpeter IMF.
Apart from the narrow interstellar Si\,{\sc ii}\,$\lambda 1526.7070$ and
C\,{\sc iv}\,$\lambda\lambda 1548.204, 1550.781$ absorption lines,
which are stronger in the distant galaxy than in the spectra
of the stars that make up the \textit{Starburst99} libraries,
the agreement between models and observations is clearly
very good. The broad blue absorption wing of the P-Cygni
profile reaches similar terminal velocities in the models
and the data, and its depth is roughly intermediate 
between those of the Galactic and Magellanic Clouds
models. The latter library of stellar spectra was built-up
by Leitherer et al. (2001) with \textit{Hubble Space Telescope}
(\textit{HST}) observations 
of stars in \emph{both} the Large and Small Magellanic Clouds 
(LMC/SMC)
and may thus correspond to an approximate metallicity
$Z_{\rm MC} \sim 0.4 Z_{\odot}$, given that in the LMC
(O/H)\,$ \simeq 0.5$\,(O/H)$_{\odot}$ 
and in the SMC (O/H)\,$ \simeq 0.25$\,(O/H)$_{\odot}$
(e.g Pagel 2003). 
By linearly interpolating between the Galactic and LMC models 
and minimising the difference from the observed spectrum,
we find
$Z_{\rm Ostars} \approx 0.6 Z_{\odot}$,
in good agreement with $Z_{\rm OBstars} = 0.5 Z_{\odot}$
deduced above from consideration of the 1425 
photospheric index.

It is interesting  that both Galactic and Magellanic Clouds
models 
seem to \textit{underpredict} the emission component
of the P-Cygni profile. 
However,  the narrow emission feature
centred near 1551.6\,\AA\ may be partly of nebular origin
(Leitherer, Calzetti, \& Martins 2002). 
Similarly, the narrow emission centred near 1533.7\,\AA\
is probably nebular Si\,{\sc ii}$^{\ast}\,\lambda 1533.4312$;
both features are missing from the synthetic spectra which
were designed to reproduce only the stellar component
of the galaxy spectrum.
The superposition of P-Cygni broad emission/absorption, 
photospheric broad absorption, narrow interstellar absorption
and narrow nebular emission attests to the complex nature of
this portion of the UV spectrum of star-forming galaxies.

In the lower panel of Figure~\ref{fig:CIV} we compare
the ESI spectra of MS\,1512-cB58 (from Pettini et al. 2002)
and the Cosmic Horseshoe in the wavelength region
1500--1570\,\AA. 
There is a startling similarity in this spectral
region between these two galaxies which are
at different redshifts (cB58 is at $z_{\rm stars} = 2.7276$)
and were selected randomly, only 
by virtue of the fact that they are highly
lensed as seen from Earth (although they are both more
luminous than the `average' galaxy at these redshifts, even
after correcting for the lensing magnifications).
Presumably, not only the chemical abundances, but also the 
young stellar populations of these two galaxies are remarkably
similar. Only the interstellar absorption lines are different,
with those in cB58 apparently stronger than in the Cosmic
Horseshoe (but see the discussion in Section~\ref{sec:ISM}).

The P-Cygni component of the C\,{\sc iv} complex 
is due to the most massive (and luminous) O stars, which drive the strongest
winds. On the other hand, a wider range of stellar spectral types,
including all stars with masses greater than about $5 M_{\odot}$, 
are thought to make up the integrated continuum light near 1550\,\AA\
(e.g. Rix et al. 2004). Thus, the contrast of the P-Cygni features
relative to the continuum is sensitive not only to metallicity,
but also to the slope and upper end cut-off of the IMF, as well
as to more subtle effects, such as differential dust extinction
among stars at the upper end of the IMF (the most massive
and short-lived stars may in principle be more reddened
than longer lived ones---Leitherer et al. 2002),
and the relative proportions of OB stars in star clusters and in the field
(Chandar et al. 2005).
Pettini et al. (2000)
showed how even relatively minor changes
to the IMF slope (or alternatively the upper mass cut-off)
result in noticeable alterations to the integrated spectrum
of a star-forming galaxy in the C\,{\sc iv} region. 
On that basis, those authors concluded that a standard
Salpeter IMF extending to $M_{\rm up} > 50 M_{\odot}$
provides the best match to the spectrum of cB58,
without the need to invoke a different IMF at high redshift,
as claimed by some.
Clearly, the same conclusion applies here to the massive
stellar population in the Cosmic Horseshoe.
More generally, the similarity evident in the lower panel
of Figure~\ref{fig:CIV} suggests that the minor effects
discussed above (differential dust extinction
and the balance between cluster and field stars) 
presumably average out in at least a subset of
star-forming galaxies at $z = 2 - 3$, when their integrated
spectra are considered.

\begin{figure*}
\vspace{-0.5cm}
\centerline{\hspace{-0.15cm}
\includegraphics[width=1.5\columnwidth,clip,angle=270]{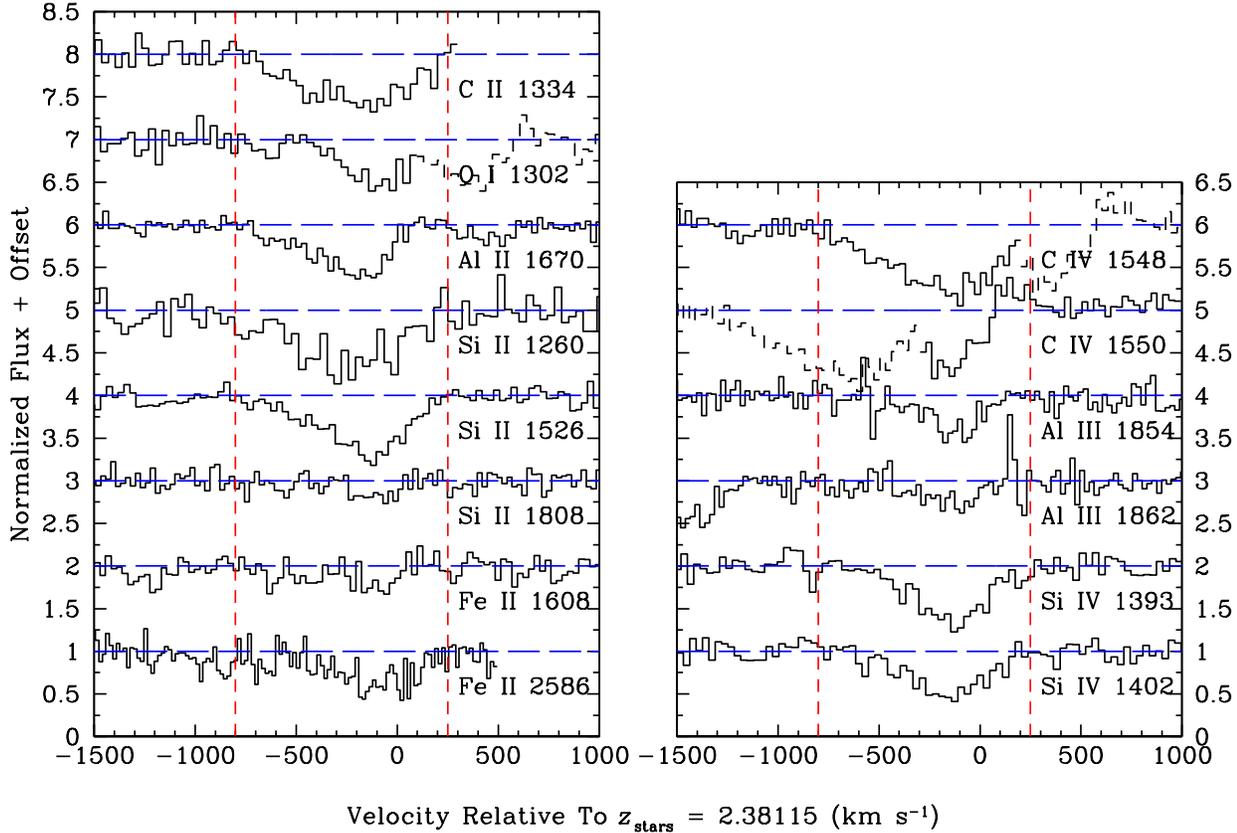}}
\vspace{-0.5cm}
\caption{Normalized  profiles  of selected interstellar absorption lines.
\textit{Left panel:} Transitions of ions which are the dominant species
of the corresponding elements in H\,{\sc i} regions. 
\textit{Right panel:} Higher ionization stages. In both panels
the vertical red dash lines indicate the velocity range over which 
the values of equivalent width listed in Table~\ref{tab:ISM} 
were generally measured.
} 
\label{fig:ISMvelocity}
\end{figure*}

Summarising the conclusions of this section,
we find that the  metallicity of the early-type stars
in the Cosmic Horseshoe is approximately half-solar,
from consideration of photospheric and wind
absorption lines. A continuous mode of star formation
with a Salpeter slope for stars more massive than
$5\,M_{\odot}$ gives a good representation of the
UV spectrum. 
The 1978 index introduced by Rix et al. (2004)
appears to be contaminated by additional features of
uncertain origin in the spectrum of the Horseshoe.
The C\,{\sc iv} region is a complex blend of stellar,
interstellar and nebular features, but the stellar
component  in the Cosmic
Horseshoe is surprisingly similar to that in cB58, 
the only other high redshift
star-forming galaxy studied  in comparable detail so far.
This may of course be a coincidence, but it may also
point to the fact that many of the parameters which
can affect the appearance of this spectral region
average out to a large extent in
the integrated UV spectrum of a whole galaxy.

\section{The Interstellar Spectrum}
\label{sec:ISM}

\subsection{Kinematics of the absorbing gas}
\label{sec:kinematics}

In the wavelength interval from 1200 to 2600\,\AA, we identified
21 interstellar absorption lines from six chemical elements in a variety of
ionization stages, from O\,{\sc i} to C\,{\sc iv}.
Relevant measurements are collected in Table~\ref{tab:ISM},
while Figure~\ref{fig:ISMvelocity} shows selected transitions chosen
to illustrate the range of elements and ion stages covered.
The interstellar lines in the Cosmic Horseshoe are broad,
with absorption 
extending smoothly from $-800$ to $+250$\,km~s$^{-1}$ 
with respect to $z_{\rm sys} = z_{\rm stars} = 2.38115$. 
Within the limits of the noise, we find no convincing 
differences between the profiles of different ions,
although the maximum velocities measured
depend on the strength of the transition,
as is the case in local starbursts (e.g. Grimes et al. 2009).
The values of $z_{\rm abs}$
listed in column (4) of Table~\ref{tab:ISM} are those
derived from the centroids of the lines (that is, the 
mean wavelength of the line
weighted by the absorption in each wavelength bin);
their mean value is $\langle z \rangle = 2.3788 \pm 0.0009$
which corresponds to a velocity offset 
$\Delta  v = -208\,$ km~s$^{-1}$ with respect to the stars.
We also measured the redshift at the peak optical
depth in the most clearly defined line profiles 
and found a mean $z_{\rm ISM} = 2.3795 \pm 0.0002$
($\Delta  v = -146\,$ km~s$^{-1}$).

This net blueshift of the interstellar lines is a common feature
of star-forming galaxies at low (e.g. Heckman et al. 2000; Martin 2005)
as well as high (e.g. Pettini et al. 2001; Shapley et al. 2003; 
Steidel et al. 2009; Vanzella et al. 2009)
redshifts. It is generally interpreted as empirical evidence for the 
existence of 
large-scale outflows of the interstellar medium driven by the
kinetic energy deposited by supernovae and the winds of 
massive stars. The values of $\Delta v$ we deduce here,
irrespective of whether one considers the mean redshift of
the line centroids or of the gas with the highest optical depth,
are typical of $z = 3$ galaxies for which Shapley et al. (2003)
derive a mean $\langle \Delta v \rangle = -150 \pm 60$\,km~s$^{-1}$.
Similarly, the widths of the strongest lines
($FWHM \simeq 400$--600\,km~s$^{-1}$)
are within the range 
$\langle FWHM \rangle = 575 \pm 150$\,km~s$^{-1}$
measured by Shapley et al. (2003) from their
composite spectrum of 811 LBGs.

\begin{table*}
 \begin{minipage}{170mm}
 \caption{\textsc{Interstellar Absorption Lines}}
   \begin{tabular}{llllccl} 
\hline 
\hline
Ion & $\lambda_{\rm lab}$$^{\rm a}$ (\AA) & $~~~~f^{\rm a}$ & $z_{\rm abs}$ & $W_0$$^{\rm b}$ (\AA)& $\sigma$$^{\rm b}$ (\AA)& Comments\\ 
\hline 
C\,{\sc ii}  	& 1334.5323	& 0.1278	          & 2.3786   &1.62  &0.16   & Blended with C\,{\sc ii}$^{\ast}\,\lambda 1335.6627$\\
C\,{\sc iv}	& 1548.204	& 0.1899	          & 2.3783   & 2.36 & 0.08  & Blended with stellar C\,{\sc iv}\,$\lambda 1549.1$; $W_0$ measured from $-800$ to $+100$\,\kms\\
	                 & 1550.781	& 0.09475	          & 2.3798  & 0.85&0.04    &  Blended with stellar C\,{\sc iv}\,$\lambda 1549.1$; $W_0$ measured from $-250$ to $+60$\,\kms\\
O\,{\sc i}     	& 1302.1685	& 0.04887	         & 2.3791  & 0.90 & 0.09 & $W_0$ measured from $-800$ to $+100$\,\kms\\
Al\,{\sc ii}        & 1670.7886	& 1.74		& 2.3779   &1.48&0.07& \\	
Al\,{\sc iii}	& 1854.7184	& 0.559	         & 2.3792  &0.93&0.10& \\
	                 & 1862.7910	& 0.278	         & 2.3782  & 0.61&0.13\\
Si\,{\sc ii}         & 1260.4221	& 1.18		& 2.3778  &1.71 & 0.18&Blended with S\,{\sc ii}\,$\lambda 1259.519$\\
	                 & 1304.3702	& 0.0863		& 2.3792  &  0.71&0.06&$W_0$ measured from $-400$ to $+25$\,km~s$^{-1}$\\
	                 & 1526.7070	& 0.133		& 2.3788  &1.81&0.06& \\
	                 & 1808.0129	& 0.00208    	& 2.3778   & 0.40&0.08&\\
Si\,{\sc iv}	& 1393.7602	& 0.513	        & 2.3795   &1.36&0.09& \\
	                 & 1402.7729	& 0.254	        & 2.3791   & 1.13&0.08&\\
Fe\,{\sc ii}	& 1608.4511	& 0.0577		& 2.3774   & 0.51&0.12\\
	                 & 2344.2139	& 0.114		& 2.3807   &0.40&0.23&\\
	                 & 2374.4612	& 0.0313		& 2.3795   &1.47&0.23&\\
	                 & 2382.7652	& 0.320		& 2.3767  & 1.25&0.18&$W_0$ measured from $-700$ to $+250$\,km~s$^{-1}$\\
	                 & 2586.6500	&  0.0691		& 2.3791  &1.85&0.13&\\
	                 & 2600.1729	& 0.239		& 2.3793  &2.25&0.28&\\
Ni\,{\sc ii}	& 1317.217	& 0.0571		& 2.3786  & 0.12 &0.05&$W_0$ measured from $-350$ to $-80$\,\kms\\
                         & 1741.5531	& 0.0427	         & 2.3799  & 0.14&0.05&$W_0$ measured from $-180$ to $-20$\,km~s$^{-1}$\\
\hline
     \label{tab:ISM}
 \end{tabular}
 $^{\rm a}$ Vacuum wavelength and $f$-values are from Morton (2003) 
with  updates by Jenkins \& Tripp (2006).\\
 $^{\rm b}$ Rest-frame equivalent width and associated error
measured over the velocity range $-800$ to +250\,\kms\ 
with respect to $z_{\rm stars}$, unless otherwise noted.
 \end{minipage}
 \end{table*}

Although with the higher resolution of our data (compared to that
normally employed to record the spectra of unlensed $z = 2 - 3$ galaxies)
the interstellar lines are fully resolved, we are still unable
to recognize clear clues in the line profiles as to the location
of the absorbing gas. The profiles are smooth and relatively featureless,
and we see no trends between, for example, velocity and degree
of ionization, which could be used to construct a kinematical model.
While the blueshifted absorption is presumably due to
outflowing material, the location and nature of the
gas moving at \emph{positive} velocities relative to the stars
remain unexplained. It is interesting to note that the absorption
lines in the ESI spectrum of MS\,1512-cB58 analysed 
by Pettini et al. (2002) also extend over the same
velocity range, from $\sim -800$ to $\sim +250$\,km~s$^{-1}$ 
relative to the systemic redshift of the stars,
although in cB58 the gas with the highest optical depth
is moving at a higher velocity $\Delta v = -255$\,km~s$^{-1}$.
The two galaxies have similar star formation rates, 
SFR\,$\approx 50 \, M_{\odot}$\,yr$^{-1}$, 
as estimated from their UV continua (see Section~\ref{sec:f_esc}).
Overall, the profiles of the interstellar absorption lines 
from the first ions appear smoother in the Horseshoe than 
in cB58, but the lower S/N of the present data makes
it more difficult to recognize distinct velocity components 
than is the case in cB58.

\subsection{Partial coverage and column densities}
\label{sec:partial_cover}


\begin{figure}
\vspace*{0.15cm}
\centerline{\hspace{-0.25cm}
\includegraphics[width=0.8\columnwidth,clip,angle=0]{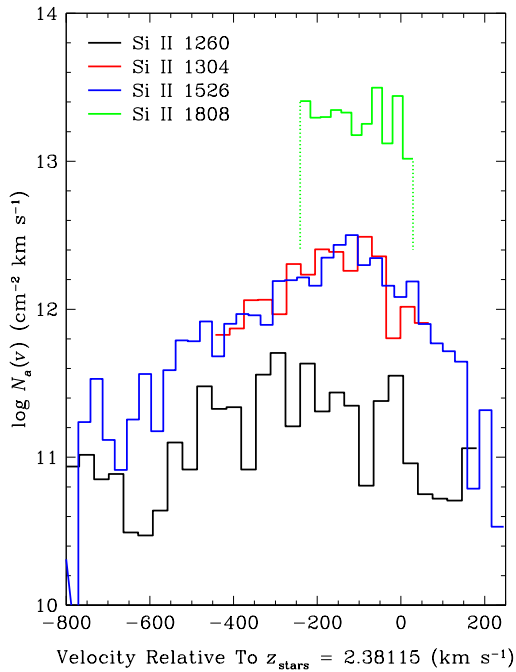}}
\caption{When analysed with the Apparent Optical Depth 
method, the four Si\,{\sc ii} transitions recorded in
our ESI spectrum of the Cosmic Horseshoe
yield values of column density which differ
systematically, in the sense expected if the 
interstellar gas does not completely cover
the stellar UV continuum
(see Section~\ref{sec:partial_cover} for further details).
}
\label{fig:AOD_SiII}
\end{figure}

One noticeable difference between the Cosmic Horseshoe and cB58 
is in the optical depth of the interstellar absorption lines 
(see Figure~\ref{fig:CIV}). Whereas in cB58 the strongest 
interstellar lines have saturated cores of zero residual intensity
(at the spectral resolution of ESI---see Figures 1 and 2 of Pettini et al. 2002), 
the same transitions never seem to reach below an optical
depth $\tau \sim 1$ in the Cosmic Horseshoe, as can be appreciated
from inspection of Figure~\ref{fig:ISMvelocity}.

One possibility is that the column densities of even the most
abundant ions are genuinely lower in the Horseshoe
compared to cB58. An alternative explanation is that the
interstellar gas does not 
completely cover the early-type stars producing
the UV continuum against which the absorption is
seen. The latter interpretation is not unlikely
in general---given the composite nature of the 
UV spectra of star-forming galaxies which are
the superposition of hundreds of thousands of
individual stellar spectra---and 
is suggested by the similarity in the residual intensity
in the cores of all the strongest absorption lines
in Figure~\ref{fig:ISMvelocity}.

We can assess quantitatively the partial coverage
hypothesis using the apparent optical depth method
of Savage \& Sembach (1991). If the absorption 
lines are resolved, as seems to be the case here,
the column density of an ion in each velocity 
bin, $N_a(v)$ [in units cm$^{-2}$ (km~s$^{-1}$)$^{-1}$],
can be deduced
directly from the optical depth in each velocity
bin, 
\begin{equation}
\tau_a(v) = - \ln \left [
I_{\rm obs}(v)/I_0(v)
\right]
\label{eq:tau1}
\end{equation}
from the relation
\begin{equation}
N_a(v) = 3.768 \times 10^{14} \cdot 
\frac{\tau_a(v)}{f \lambda}
\label{eq:tau2}
\end{equation}
where $f$ measures the strength of the transition
at wavelength $\lambda$  (\AA), and
$I_{\rm obs}$ and $I_0$ denote the relative
intensities in the line and in the continuum respectively.
If multiple absorption lines arising from the same 
ground state of an ion but with different values of
the product $f \lambda$
are available, partial coverage would manifest 
itself as a mismatch between the values of 
column density deduced from each transition, 
the lines with smaller values of $f \lambda$ giving 
systematically higher values of $N_a(v)$.

Among the ions covered by the ESI spectrum of
the Horseshoe, Si\,{\sc ii} is the one best suited
to this analysis, with four transitions spanning
a range of $\sim 400$ in $f \lambda$
(see Table~\ref{tab:ISM}) and falling at
wavelengths where the S/N of the data
is highest.  
As can be seen from Figure~\ref{fig:AOD_SiII},
the run of $N_a(v)$ vs. $v$ for the four
Si\,{\sc ii} lines does indeed show
the systematic differences 
expected if the absorbing gas
does not completely cover the background
stellar continuum. 
In the line core, the apparent optical depth
of the weakest absorption
line, Si\,{\sc ii}\,$\lambda 1808.0129$, 
implies values of column density \emph{two orders of
magnitude} higher than those deduced from the 
strongest transition, Si\,{\sc ii}\,$\lambda 1260.4221$. 
The profiles of
Si\,{\sc ii}\,$\lambda 1304.3702$ and $\lambda 1526.7070$,
with very similar values of $f \lambda$ intermediate
between those of the other two lines, indicate
that partial coverage extends well beyond the 
line cores and probably applies to the full velocity
range spanned by the absorption.

We also cover several Fe\,{\sc ii} transitions
(see Table~\ref{tab:ISM}), but their $f \lambda$
values range over a factor of only $\sim 10$ and
most of them are redshifted to far-red wavelengths, 
where the S/N of our data is lower. 
Nevertheless, the profiles of the Fe\,{\sc ii} lines
are consistent with the conclusions drawn from 
Figure~\ref{fig:AOD_SiII}, in showing a systematic
trend of increasing $N_a(v)$  with decreasing $f \lambda$.
Turning to the high ionization lines  in the right-hand
panel of Figure~\ref{fig:ISMvelocity}, we found that the
weaker member of the Si\,{\sc iv} doublet, $\lambda 1402.7729$,
also gives systematically higher values of $N_a(v)$
than $\lambda 1393.7602$, even though the $f \lambda$
values of the two lines 
differ by only a factor of $2$, while the evidence
is less clear-cut in the case of the Al\,{\sc iii} doublet
which is recorded at lower S/N. The C\,{\sc iv} lines
are difficult to interpret in this context because
they are  blended with each other and with the 
stellar P-Cygni profile, as discussed above 
(Section ~\ref{section:stellarwind}).

\begin{figure}
\vspace*{0.25cm}
\centerline{\hspace{-0.175cm}
\includegraphics[width=0.725\columnwidth,clip,angle=270]{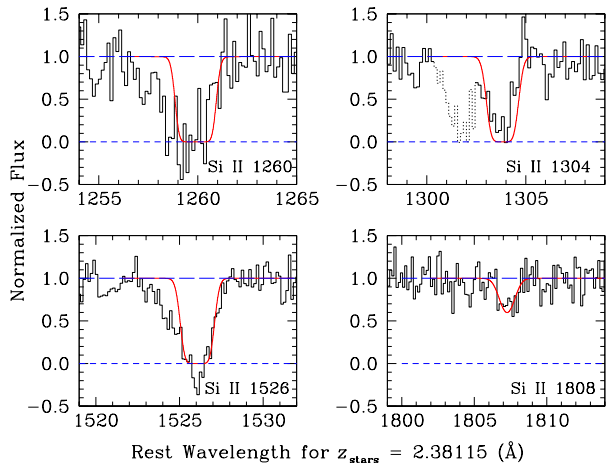}}
\vspace{0.25cm}
\caption{\textit{Black histograms:} Normalized profiles of the
four Si\,{\sc ii}  absorption lines after subtracting 40\% of the 
continuum level, under the assumption that the gas only
covers 60\% of the stellar light. 
\textit{Red continuous lines:} Theoretical absorption line
profiles generated with the values of $N$(Si\,{\sc ii}) and
velocity dispersion parameter $b$
implied by the weakest transition, $\lambda 1808.0129$,
which is unsaturated. 
The absorption line shown by the short-dash histogram 
in the top right-hand panel
is O\,{\sc i}\,$\lambda 1302.1685$.
}
\label{fig:SiII_adjusted}
\end{figure}

Given our lack of knowledge of the relative configuration
of interstellar gas and stars, it is quite possible, 
or even likely (e.g. Martin \& Bouch{\'e} 2009), 
that the covering 
factor varies with velocity and degree of ionization of the gas.
However, the S/N of our data is insufficient to establish
the magnitude of such variations. Instead, we adopted the
simplest assumption that ``one size fits all'' and corrected
the absorption line profiles by re-adjusting the zero level
by the same fractional amount of the continuum at all
wavelengths.  Inspection of the spectrum shows that
even the strongest absorption lines, which are normally
saturated in the spectra of Galactic stars and of nearby
starbursts, all seem to have a minimum residual intensity
$I_{\lambda}/I_0 \sim 0.4$ (see Figure~\ref{fig:ISMvelocity}).
With the simplistic assumption that the interstellar gas only 
covers 60\% of the stellar light, and that the remaining 40\%
of the UV continuum sees no absorption in our direction,
the adjusted profiles of the four Si\,{\sc ii} lines become
internally consistent, as shown in Figure~\ref{fig:SiII_adjusted}.

Specifically, after subtracting 40\% of the stellar continuum
and renormalizing the line profiles, we deduced the column
density of Si\,{\sc ii}, 
$\log N$(Si\,{\sc ii})/cm$^{-2} = 16.00$,
by integrating eq.~(\ref{eq:tau2})
over the velocity range spanned by the weakest
Si\,{\sc ii} line, $\lambda 1808.0129$, which is unsaturated
even after the readjustment of the zero level.
We then used the absorption line fitting 
software {\sc xvoigt} (Mar \& Bailey 1995) 
to fit the profile of Si\,{\sc ii}\,$\lambda 1808.0129$
with this column density and deduced the
value of the velocity dispersion parameter, $b = 105$\,km~s$^{-1}$,
which best reproduces the width of this line.
In the next step, we used this pair 
of values of $N$(Si\,{\sc ii}) and $b$
to generate {\sc xvoigt} profiles
of the other three Si\,{\sc ii} absorption lines. 
As can be seen fom Figure~\ref{fig:SiII_adjusted}, these computed
profiles provide satisfactory fits to the main absorption component
of the three stronger Si\,{\sc ii} transitions, given the noise in the data.
(We have not attempted to fit the blue wing of the three
lines because this absorption is
too weak to be detected in Si\,{\sc ii}\,$\lambda 1808.0129$,
but its contribution to the value of $N_{\rm TOT}$(Si\,{\sc ii}) is
only a small fraction of the total, unless the covering factor of the
high velocity gas is much smaller than 60\%).

Table~\ref{table:N} lists the values of ion column density 
deduced by integrating eq.~(\ref{eq:tau2}) over the velocities
spanned by those absorption lines that remain unsaturated
after the 40\% adjustment of the zero level. Errors in the
column density, $\delta N$, were determined by adding/subtracting the 
$1 \sigma$ error spectrum to/from the line profiles and
recalculating the column density; in cases where the addition
of the noise spectrum results in the line becoming saturated
no corresponding error is given in Table~\ref{table:N}.
Again, we see from the Table that there is good internal 
agreement between the values of $N$ implied by different
transitions of the same ion, after correcting for the 40\%
unabsorbed stellar continuum.

\begin{table}
 \begin{minipage}{152mm}
 \caption{\textsc{Ion Column Densities}}
   \begin{tabular}{lllcl} 
\hline 
\hline
Ion & $\lambda_{\rm lab}$ (\AA) & $~~~f$ & $\log N$/cm$^{-2}$& $\log \delta N$/cm$^{-2}$ \\ 
\hline 
Al\,{\sc iii} 	& 1854.7184	        & 0.559	&  14.22    &$-0.06$, ...\\
	                 & 1862.7910                & 0.278	& 14.26     &$-0.07$, +0.14\\
Si\,{\sc ii}          & 1808.0129              & 0.00208   & 16.00     & $-0.06$, +0.07 \\
Si\,{\sc iv}  	&1402.7729	       & 0.254	&14.92      &$-0.06$, ...\\
Fe\,{\sc ii}        & 1608.4511	       & 0.0577	& 14.94     &$-0.10$, +0.11\\
                         & 2586.6500              &  0.0691    & 15.12     & $-0.06$, ...\\
Ni\,{\sc ii}        & 1317.217               & 0.0571     &  14.22    & $-0.24$, +0.20\\
                         & 1741.5531             & 0.0427      & 14.40     & $-0.21$, +0.19\\
\hline
     \label{table:N}
 \end{tabular}
 \end{minipage}
 \end{table}

The interstellar column densities 
of the first ions in the Cosmic
Horseshoe are surprisingly 
similar to those measured in cB58 by Pettini et al. (2002).
For the transitions in common between
the two studies, we have:
$\log N$/cm$^{-2}$~(Horseshoe/cB58)\,$ = 16.00/15.99$ for
Si\,{\sc ii}\,$\lambda 1808.0129$;
$ 14.94/15.17$ for Fe\,{\sc ii}\,$\lambda 1608.4511$ ;
$ 14.22/14.32$ for Ni\,{\sc ii}\,$\lambda 1317.217$;
and $ 14.40/\leq 14.45$ for
Ni\,{\sc ii}\,$\lambda 1741.5531$.
Although the metallicities reached by the two galaxies
are similar, as discussed above (Section~\ref{sec:stellar_spectrum}),
there was no reason to expect that the column densities
of interstellar gas entrained by the outflows---nor the
ionization of this gas---should be so similar.
Furthermore, even the ratios of elements that are
depleted onto dust by different degrees, or are synthesised 
by stars of different masses, are similar between these
two galaxies. Thus, the iron-peak and easily depleted
elements Fe and Ni are underabundant (in their first ions)
compared with the undepleted alpha-capture element Si by about one
order of magnitude, as is the case in cB58 (see Figure 8
of Pettini et al. 2002).
All of this may just be a coincidence, of course, or may point
to a more fundamental uniformity in the conditions of
the gas outflowing from high-redshift star-forming galaxies, or at least
a subset of such galaxies. 

Given the similarity in the column densities of the first ions
and in the metallicity of the two galaxies, it follows that
we also expect similar values of the column density of
hydrogen $N$(H). 
Indeed, if
the majority of the first ions
arises from H\,{\sc i} regions, we can deduce values 
of $N$(H\,{\sc i}) from the column densities 
in Table~\ref{table:N} and the assumption that
$Z_{\rm ISM} = Z_{\rm OBstars} = 0.5 Z_{\odot}$.
This leads to 
$\log N$(H\,{\sc i})/cm$^{-2} = 20.79$
using Si\,{\sc ii} as the reference ion,
indeed very close to 
$\log N$(H\,{\sc i})/cm$^{-2} = 20.85$
deduced by Pettini et al. (2002) from the 
damped profile of the \lya\ line in cB58.
As we shall see in a moment, however,
there are striking differences between the 
profiles of the \lya\ line observed in the
two galaxies. Whereas in the case of cB58 
we see a clear damped absorption profile
with a weak superposed emission component,
\lya\ is in emission in the Cosmic Horseshoe.

\section{The Lyman alpha Line}
\label{sec:lya}

\subsection{Lyman $\alpha$ line morphology}

The top panel of Figure~\ref{fig:LyAlphaVel} shows the spectral
region encompassing the \lya\ line in the Cosmic Horseshoe,
plotted in arbitrary flux-density units 
vs. velocity relative to the systemic
redshift of the galaxy.
We interpret the complex morphology as 
consisting of two main emission 
components: a narrow, symmetric component 
which is barely resolved
(FWHM\,$\simlt 50$\,km~s$^{-1}$ after deconvolution
with the ESI spectral resolution of  FWHM\,$= 75$\,km~s$^{-1}$)
centred at $v = + 115$\,km~s$^{-1}$, and a broader,
asymmetric component which peaks at $v = +275$\,km~s$^{-1}$
but extends to $v \simeq +700$\,km~s$^{-1}$.
The first, narrow component has a peak flux $\sim 1.5$ times 
higher than that of the second.
There may be a third, weak component at negative
velocities, centred near $v \sim -80$\,km~s$^{-1}$,
but its identification is uncertain, given the blending
with absorption from the \lya\ forest.
The total flux in the line, without
correcting for gravitational lensing and reddening
(these corrections will be applied later---see Section~\ref{sec:f_esc})
is $F({\rm Ly}\alpha) = (4.5 \pm 0.2) \times 10^{-16}$\,erg~s$^{-1}$~cm$^{-2}$
where the error reflects both the random noise in the data
and the uncertainty in the stellar continuum (measured
redward of \lya), combined in quadrature. 
In our assumed cosmology, 
the measured flux corresponds to a \lya\ luminosity 
$L({\rm Ly}\alpha) = (2.0 \pm 0.1) \times 10^{43}$\,erg~s$^{-1}$.
The line equivalent width
is $W_0({\rm Ly}\alpha) = (11 \pm 0.2 \pm 2.5)$\,\AA,
(switching sign convention, compared to Table~\ref{tab:ISM},
to indicate equivalent widths of emission lines 
as positive and those of absorption lines as negative);
here we have quoted separately the random and systematic errors
respectively to emphasize that the main uncertainty in the value
of $W_0({\rm Ly}\alpha)$ arises from the placement of the 
underlying stellar continuum.
Although the net profile is in emission, the 
value of $W_0({\rm Ly}\alpha)$ we measure is
below the threshold $W_0({\rm Ly}\alpha) \geq 20$\,\AA\
generally adopted to define the so-called \lya\ emitters
(e.g.  Hu et al. 1998; Rhoads et al. 2000; Shapley et al. 2003).
Nevertheless, the profile exhibits the broad characteristics
that are typical of \lya\ emission from high-$z$ galaxies:
an abrupt blue edge and an
extended red wing (e.g. Tapken et al. 2007); with the
high resolution and S/N of our data we can now
examine its characteristics in greater detail than
is normally possible.

\subsection{Comparison to radiative transfer models}
\label{sec:rad_trans}

\begin{figure}
\vspace{-1.45cm}
\centerline{\hspace{0.15cm}
\includegraphics[width=1.08\columnwidth,clip,angle=0]{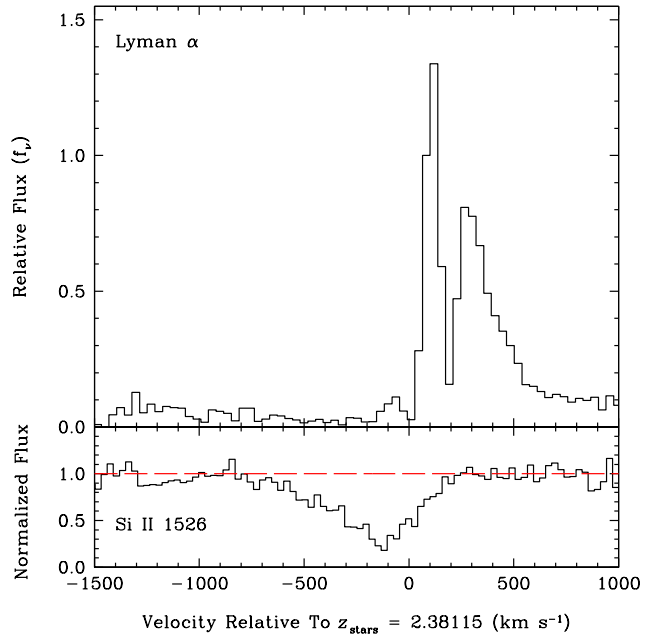}}
\vspace{-2cm}
\caption{\lya\ emission and Si\,{\sc ii} absorption lines plotted
on a common velocity scale. The former mirrors the latter in
velocity space.}
\label{fig:LyAlphaVel}
\end{figure}

The \lya\ emission line emerging from star-forming galaxies
has been the subject of many studies over the last forty years,
since the seminal paper by Partridge \& Peebles (1967).
Among more recent investigations, those by the Geneva group 
(see, for example, Schaerer 2007 which also includes a 
comprehensive set of references to earlier work) 
have focussed in particular on the 
radiation transfer of \lya\ photons in an expanding medium,
in an attempt to reproduce the variety of \lya\ profiles seen
in high-redshift galaxies, where large-scale outflows are the norm.
Having established the existence of such outflows in the 
Cosmic Horseshoe (Section~\ref{sec:kinematics}),
we can use the models by Verhamme et al. (2006)
to interpret the \lya\ profile in Figure~\ref{fig:LyAlphaVel}.

The basic idea is that the \lya\ photons emitted from 
a central region of star-formation, where the early-type stars
and their H\,{\sc ii} regions are located, suffer multiple 
scatterings before they can escape the nebula. 
Such scatterings not only redistribute the photons in frequency, 
but also convert a fraction of them into infrared photons
if any dust is present.
If the scattering medium is expanding, the net effect is
a reduction of the \lya\ luminosity (measured by the fraction,
$f_{\rm esc}$, of photons which manage to escape)
and an overall shift of the wavelength distribution 
of the escaping photons to longer wavelengths,
as blue-shifted photons are absorbed by the H\,{\sc i} 
gas in front of the stars.
The profile of the emergent \lya\ line 
(for a given star formation rate which
determines the intrinsic \lya\ luminosity) depends on 
a combination of parameters, principally the 
expansion velocity of the medium, $v_{\rm exp}$;
its internal velocity dispersion, normally measured by
the parameter $b = \sqrt{2} \sigma$;
the column density of neutral hydrogen, $N$(H\,{\sc i});
and the column of dust, as measured by the colour excess
$E(B-V)$.

The \lya\ line in
the Cosmic Horseshoe matches well 
some of the characteristics of the 
theoretical profiles generated by the Verhamme et al. (2006)
models for the simplest geometrical configuration:
a spherical shell expanding with uniform velocity.
Indeed, as can be appreciated from Figure~\ref{fig:LyAlphaVel}
where the \lya\ emission line and the Si\,{\sc ii}\,$\lambda 1526.7070$
absorption line are plotted on a common velocity scale,
the red wing of the former almost mirrors the blue wing of
the latter, both extending to $| v | \simeq 700$--800\,km~s$^{-1}$.
A prediction of the models is that there should be a
maximum in the \lya\ emission at a positive velocity 
(relative to the stars) $v = -2 v_{\rm exp}$. 
In our case, the velocity of the longer-wavelength peak,
$v = +275$\,km~s$^{-1}$ is indeed at about twice
$|v| = 146\,$ km~s$^{-1}$ which we measured
in Section~\ref{sec:kinematics} for the gas with the highest
optical depth in absorption.
Referring to Figure~12 of Verhamme et al. (2006),
these are photons which have undergone 
just one backscattering from the receding portion 
of the expanding shell into
our line of sight---with this frequency shift they 
can travel unabsorbed through the approaching
(i.e. blue-shifted) part of the shell.
The long tail to $v \simeq 700$\,km~s$^{-1}$ is made
up of photons which have undergone two or more
backscatterings, while the narrow component 
centred at $v = + 115$\,km~s$^{-1}$
consists of photons which have been 
multiply scattered from the
approaching part of the shell.

The balance between the photons  escaping from
the approaching and receding portions of the 
shell---reflected in the relative flux of the two 
emission peaks---is a function of the column
density of neutral hydrogen through the shell: 
as $N$(H\,{\sc i})
increases, the more redshifted peak of emission
dominates over the lower redshift one.
Out of the illustrative examples considered
by Verhamme et al. (2006) in their Figure~16,
the case with
$N$(H\,{\sc i})\,$= 7 \times 10^{19}$\,cm$^{-2}$
matches our observed profile closely.

While this correspondence between simulated and 
observed profiles is impressive, other aspects of the 
models are less satisfactory. 
One of the difficulties in relating the Verhamme et al. (2006)
models to our observations of the Cosmic Horseshoe
stems from the fact that the absorption profiles of the interstellar lines
do not fit the simple picture of an expanding shell with a 
well-defined velocity. With absorption extending
over a thousand km~s$^{-1}$, and without any information
which would allow us to relate the velocity of the gas 
to its location within the galaxy, there is no obvious justification for 
equating the velocity of the gas with the highest optical depth,
$v = -146\,$ km~s$^{-1}$, to the expansion
velocity of the idealised shell.
A further difficulty is that in the Verhamme et al. (2006) models
there is also a strong dependence of the line profile on the
velocity dispersion of the gas within the expanding shell.
The clear distinction between the two redshifted peaks of
emission requires not only low column densities of
neutral gas [$N$(H\,{\sc i})\,$\simlt 5 \times 10^{20}$\,cm$^{-2}$],
but also relatively quiescent gas within the expanding shell
($|v_{\rm exp}|/b \simgt 5$).
With the values we measure from the interstellar lines
in the Horseshoe, $|v_{\rm exp}|/b = 145/105 \sim 1$,
the profiles predicted by Verhamme et al. (2006)
would look drastically different (see their Figure~15)
and the two emission components would merge into
one. The problem is exacerbated by the presence of gas 
at velocities as high as $v \simeq -800$\,km~s$^{-1}$
(see Figure~\ref{fig:ISMvelocity})
which is not included in the models.
A final point  is that the
models considered by Verhamme et al. (2006)
generally assume unity covering factor of
the central source of \lya\ photons by the outflowing
gas, whereas in the Cosmic Horseshoe the interstellar
absorption lines appear to cover only $\sim 60$\% of
the stellar UV continuum (Section~\ref{sec:partial_cover}).

Presumably, the real geometrical configuration of stars,
H\,{\sc ii} regions (which recall are 
at the same velocity as the stars---see Table~\ref{table:systemic_z}),
and outflowing ISM is more complicated than in the simple
picture considered (for general illustrative purposes)
by Verhamme et al. (2006). 
Treatments of \lya\ radiation
transfer in more complex situations have been examined  
in the literature, but generally with fewer
specific realizations that can be compared directly
to our observations of the \lya\ line in the Cosmic Horseshoe.
For example, Hansen \& Oh (2006) considered the
effects of a multiphase gas outflow, either in a shell with 
holes or in an ensemble of gas clumps. It is interesting 
that in their simulations the former scenario can still
give a double-peak \lya\ emission line, 
whereas the latter tends to
generate more amorphous profiles. 
A further point of note is that in such multiphase 
outflows \lya\ photons can escape even when
the intervening column density of neutral hydrogen is high.
It would obviously be of considerable
interest to test whether a model tailored to the observed
properties of the Cosmic Horseshoe can reproduce
its \lya\ profile.


\subsection{Escape fraction of \lya\ photons}
\label{sec:f_esc}

Before concluding this section,
we calculate the escape fraction of \lya\
photons by comparing the observed \lya\
luminosity with that of the 
UV continuum. In carrying out such  a comparison,
we convert the luminosities to star formation rates (SFR)
using the conversions proposed by Kennicutt (1998),
divided by a factor of 1.8 to account for the lower proportion
of low mass stars in the Chabrier (2003) IMF
relative to the standard Salpeter (1955) IMF adopted by Kennicutt.
We also correct for gravitational lensing magnification,
although our conclusion is not sensitive to this factor, 
as we are comparing luminosities over the same regions
of the Cosmic Horseshoe. 
In the lensing model by Dye et al. (2008),
the magnification factor of  
the whole Einstein ring is 24;  
in Section~\ref{sec:observations} we calculated that
the ESI slit captured a fraction $0.25 \pm 0.05$ 
of the total light. In the following, we do not carry
the uncertainty in this estimate through the calculation,
as it is the same for the \lya\ line and the UV continuum,
and we adopt a magnification factor of $24 \times 0.25 = 6$
for the portion of the Cosmic Horseshoe
recorded with ESI.
The comparison between
SFR(Ly$\alpha$) and SFR(UV)
is sensitive to the reddening correction, since the two
estimates are based on features at different wavelengths.
We adopt $E(B-V) = 0.15$, as  determined 
for the same two knots of the Horseshoe 
by Hainline et al. (2009---see
their discussion of the derivation of this value),
together with the reddening curve of Calzetti et al. (2000).

From our ESI spectrum we measure
a UV continuum luminosity at 1700\,\AA\
$L_{\nu}(1700\,{\rm \AA}) = 1.1 \times 10^{30}$\,erg~s$^{-1}$~Hz$^{-1}$ 
which, using Kennicutt's (1998) conversion
\begin{equation}
{\rm SFR(UV)} (M_{\odot}~{\rm yr}^{-1})
= 1.4 \times 10^{-28} L_{\nu} ~ ({\rm erg~s^{-1}~Hz^{-1})} ,
\label{eq:SFR_UV1}
\end{equation}
gives 
\begin{equation}
{\rm SFR(UV)}\,
= 159 \times 1/1.8 \times 3.8 \times 1/6 =  56 M_{\odot}~{\rm yr}^{-1}
\label{eq:SFR_UV2}
\end{equation}
with correction factors for the IMF, 
reddening, and magnification respectively.\footnote{
Recently, Leitherer (2008) re-examined Kennicutt's 
calibrations as given here in eqs.~(\ref{eq:SFR_UV1}) and (\ref{eq:SFR_Lya1})
in the light of new \textit{Starburst99} models which include the 
effects of stellar rotation. While these models are still 
at an exploratory stage, the indications are that,
at solar metallicities,  they
may result in $\sim 20$\% reductions 
of the values of SFR deduced from
both the UV continuum and H$\alpha$. 
At sub-solar metallicities, however,
SFR(H$\alpha$) is further reduced relative to SFR(UV),
reflecting the harder ionizing spectrum of metal-poor massive
stars. Such effects may partly explain why we
deduce SFR(H$\alpha) > $\,SFR(UV) in the Cosmic Horseshoe.}

For comparison,
Hainline et al. (2009) deduced 
SFR(H$\alpha) = 113 \pm 17 M_{\odot}$~yr$^{-1}$
from the H$\alpha$ luminosity measured
with NIRSPEC,
but the comparison is complicated 
by the uncertainties in the respective flux calibrations.
In any case, in galaxies at $z = 2 - 3$, there is an 
inherent dispersion  (of about this magnitude)
in star formation measurements
from H$\alpha$ emission and UV continuum (e.g. Pettini et al.  2001; Erb 
et al. 2006c) which 
can have a number of causes, apart from random errors.
First, emission lines and UV continuum
do not sample the same stellar populations:
the luminosity of H$\alpha$ is due primarily to the most massive
stars, with shorter lifetimes than the wider range of stellar
masses ($M \simgt 5 M_{\odot}$) whose integrated light makes
up the continuum at 1700\,\AA.
Second, the conversion from H$\alpha$
(and Ly$\alpha$) luminosity to SFR implicitly assumes that
all the Lyman continuum (LyC) photons are absorbed within the
H\,{\sc ii} region and reprocessed into emission lines---a
situation which is often referred to as a `radiation-bounded nebula'.
Of course, if a fraction of LyC photons escape unabsorbed
from the H\,{\sc ii} region, SFR(H$\alpha$) and SFR(Ly$\alpha$) 
will be systematically lower than SFR(UV).

Turning to \lya, we can use Kennicutt's (1998)
calibration of SFR(H$\alpha$) and case B recombination to
deduce
\begin{equation}
{\rm SFR(Ly}\alpha)  (M_{\odot}~{\rm yr}^{-1})
=  9.1 \times 10^{-43} L({\rm Ly}\alpha) ~ ({\rm erg~s^{-1})}
\label{eq:SFR_Lya1}
\end{equation}
which, together with  our measured
$L$(Ly$\alpha)  
= 2.0 \times 10^{43}$\,erg~s$^{-1}$,
gives
\begin{equation}
{\rm SFR(Ly}\alpha) = 
18 \times 1/1.8 \times 5.3 \times  1/6 = 8.9 M_{\odot}~{\rm yr}^{-1}
\label{eq:SFR_Lya2}
\end{equation}
where, as in eq.~(\ref{eq:SFR_UV2}), the correction factors account for
the IMF, dust extinction, and magnification respectively.

By comparing the three estimates of SFR,
we deduce $f_{\rm esc}^{\rm Ly\alpha} \simeq 0.16$--0.08,
depending on whether we compare SFR(Ly$\alpha$) to
SFR(UV) or SFR(H$\alpha$), respectively. 
The former value is independent
of the absolute flux calibration
but assumes that, intrinsically (i.e. before the
\lya\ line is quenched by resonant scattering), 
SFR(UV)\,$ \equiv $\,SFR(Ly$\alpha$).
Conversely, the latter value assumes no errors in 
the relative flux calibrations of the NIRSPEC and ESI data,
and in the calculation of the fractions 
of the total light from the Horseshoe 
captured by the two spectrographs.
However, either estimate of $f_{\rm esc}^{\rm Ly\alpha}$
is plausible, given the wide range found in other \lya\ emitters 
(e.g. Verhamme et al. 2008; Pentericci et al. 2008).
It must also be remembered that these values
of  $f_{\rm esc}^{\rm Ly\alpha}$ apply to
our viewing angle and to the area of the galaxy
covered by the entrance slit of the spectrographs;
the escape fraction could be higher in other
directions (Neufeld 1991) and over larger areas
(Saito et al. 2006).

Concluding this section, 
models of \lya\ emission resonantly scattered
by an expanding medium need
further improvement in order to reproduce
simultaneously the observed properties
of the \lya\ emission line and of the interstellar
absorption lines in the Cosmic Horseshoe.
Nevertheless, interpreting the \lya\ line in
terms of these models has given us some insight
into the physical conditions of the gas.
The main inference one would draw from 
the models of Verhamme et al. (2006) is that
the column densities of neutral hydrogen and 
dust need to be relatively 
low---$N$(H\,{\sc i})\,$ \simeq  7 \times 10^{19}$\,cm$^{-2}$
and $E(B-V) \simeq 0.1$---in order for the line 
to exhibit two distinct
emission peaks at the velocities observed.
On the other hand, recall
that in Section~\ref{sec:partial_cover}
we concluded that 
$\log N$(H\,{\sc i})/cm$^{-2} \simeq 20.8$
\emph{if} the first ions arise primarily in H\,{\sc i} gas,
as is the case in the Milky Way, damped \lya\ systems,
and the gravitationally lensed LBG cB58.
The order of magnitude difference between
these two estimates of the column density of gas
can be reconciled if most of the gas in front of the stars
in the Cosmic Horseshoe is ionized, with a neutral fraction
of only about 10\%. 
The profile of the
O\,{\sc i}\,$\lambda 1302.1685$ absorption line, 
which traces exclusively neutral gas, 
is consistent with this conclusion.


\begin{figure}
\vspace*{-1.4cm}
\centerline{\hspace{-0.15cm}
\includegraphics[width=1.1\columnwidth,clip,angle=0]{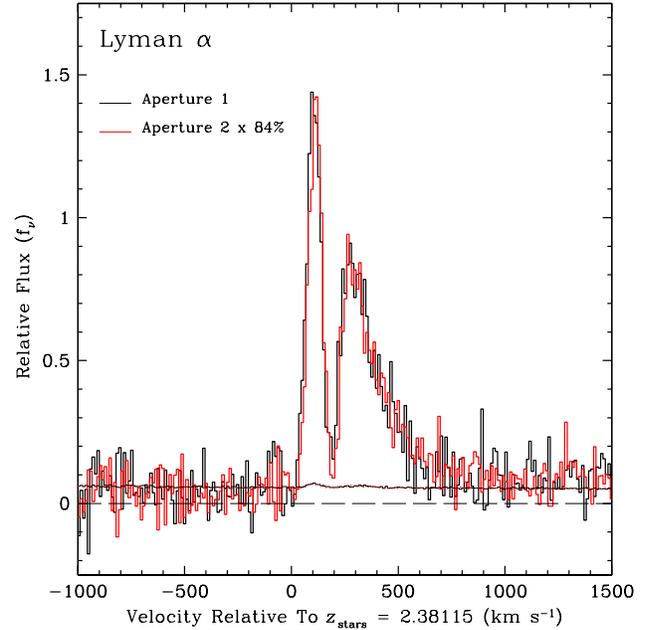}}
\vspace{-2cm}
\caption{After scaling the flux in Aperture 2 
(see Figure~\ref{fig:horseimage}) by a factor of 0.84,
the \lya\ profiles recorded from the two bright knots
of emission in the Cosmic Horseshoe
are indistinguishable from each other
at the S/N of the present observations.
The corresponding error spectra are also plotted 
(thin line near the zero level).
}
\label{fig:LyAlphaAps}
\end{figure}

\begin{table*}
\centering
 \begin{minipage}[c]{105mm}
 \caption{\textsc{C iii] Emission Lines}}
 \begin{tabular}{@{}lllcc} 
\hline 
\hline
 \multicolumn{1}{c}{Line}
  & \multicolumn{1}{c}{$z$} 
  & \multicolumn{1}{c}{$\sigma^{\rm a}$}
  & \multicolumn{1}{c}{$F^{\rm b}$}
  & \multicolumn{1}{c}{$L^{\rm b}$}\\
  \multicolumn{1}{c}{ } 
  & \multicolumn{1}{c}{ }
  & \multicolumn{1}{c}{(km~s$^{-1}$)}
  & \multicolumn{1}{c}{($10^{-18}$erg~s$^{-1}$~cm$^{-2}$)}
  & \multicolumn{1}{c}{($10^{41}$\,erg~s$^{-1}$)}\\
\hline 
1906.683 & $2.38103 \pm 0.00014$  & $62 \pm 7$  & $9.9 \pm 1.3$ & $1.29 \pm 0.16$\\
1908.734 & $2.38127 \pm 0.00017$  & $62 \pm 7$  & $8.8 \pm 1.3$ & $1.14 \pm 0.16$\\
\hline 
     \label{tab:CIII}
 \end{tabular}

$^{\rm a}$ After correcting for the instrumental resolution 
($\sigma_{\rm instr} = 32$\,km~s$^{-1}$), and constrained to
be the same for both lines.\\
$^{\rm b}$ Corrected for reddening and lensing magnification.\\
 \end{minipage}
 \end{table*}

Given the sensitivity of the profile of the \lya\ line
to the geometrical and physical
properties of the ambient interstellar medium,
one may expect to see variations between 
even slightly different sightlines
through the galaxy. In principle,
the lensing magnification of the source offers 
a good opportunity to recognize such small-scale
inhomogeneities. And yet no such variations are
seen in the Cosmic Horseshoe. 
Recall (Section~\ref{sec:observations} and
Figure~\ref{fig:horseimage}) that we obtained 
separate spectra for the two brightest knots
in the Einstein ring, which we labelled 
`Aperture 1' and `Aperture 2'.
After scaling the latter by a factor of 0.84
(derived empirically and found to be the same 
for the \lya\ line and the UV continuum),
the profiles of the \lya\ line in the two apertures are
identical within the limits of the noise,
as can be appreciated from inspection
of Figure~\ref{fig:LyAlphaAps}.
Evidently, the same region of the source is lensed into
the two knots of the Cosmic Horseshoe.
This is all the more puzzling, given that Hainline
et al. (2009) found that the ratio
[N\,{\sc ii}]\,$\lambda 6584$/H$\alpha$
differs by almost a factor of 2 
(at the $\sim 4 \sigma$ significance level)
between the two Apertures. 
The [N\,{\sc ii}]/H$\alpha$ ratio responds to
changes in metallicity and ionization parameter
(e.g. Pettini \& Pagel 2004; Kobulnicky \& Kewley 2004)
which presumably are not the same in the regions of the 
galaxy lensed into Apertures 1 and 2 within
the Einstein ring (knots `A' and `D' of Belokurov et al. 2007).
Evidently,  such variations do not have an impact on the
\lya\ emission morphology.

\section{Other Emission Lines}
\label{sec:CIII}

\begin{figure}
\vspace*{-2.75cm}
\centerline{\hspace{0.5cm}
\includegraphics[width=1.1\columnwidth,clip,angle=0]{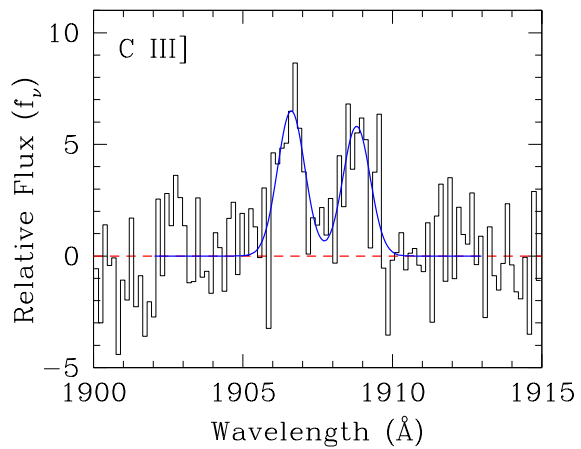}}
\vspace{-3.25cm}
\caption{\textit{Histogram:} Portion of the ESI spectrum of the Cosmic
Horseshoe in the region of the 
C\,{\sc iii}]\,$\lambda \lambda 1906.683, 1908.734$ 
nebular emission lines, after subtracting 
the stellar continuum.
\textit{Blue continuous line:} Gaussian fits to the emission lines;
see Table~\ref{tab:CIII} and Section~\ref{sec:CIII} for details.
}
\label{fig:CIII}
\end{figure}

The nebular C\,{\sc iii}]\,$\lambda\lambda 1906.683, 1908.734$
doublet is clearly resolved in our ESI spectrum of the Cosmic Horseshoe
(see Figure~\ref{fig:CIII}).
By fitting Gaussian profiles to the two lines, we deduced the
parameters listed in Table~\ref{tab:CIII}.
The mean redshift, $z_{\rm C\,\textsc {iii}]} = 2.38115$,
is in excellent agreement with the redshift of the
OB stars, as expected (Section~\ref{sec:z_sys}),
and differs by only $-7$\,km~s$^{-1}$ from the weighted mean
$\langle  z_{\rm H \alpha} \rangle = 2.38123$ of the
two apertures (see Figure~\ref{fig:horseimage})
measured separately 
by Hainline et al.  (2009) from
their NIRSPEC spectra.
The good match attests to the mutual consistency of the ESI
and NIRSPEC wavelength scales.
The line widths are rather uncertain due to the limited
S/N;  with the prior condition that the width should be the same 
for both lines, we obtain 
$\sigma_{\rm C\,\textsc{iii}]}= 62 \pm 7$\,km~s$^{-1}$
(this is the value used in the fits shown in Figure~\ref{fig:CIII}),
in very good agreement---again, as expected---with
the weighted mean
$\langle  \sigma_{\rm H \alpha} \rangle = 65$\,km~s$^{-1}$
of the two apertures reported by Hainline et al. (2009).

As is well known, the ratio of these two C\,{\sc iii}]
lines is a function of
the electron density, varying from 
$F(1906)/F(1908) = 1.5$ to  $\sim 0.8$ in the range 
$n(e) = 100$--30\,000\,cm$^{-3}$.
Our measured $F(1906)/F(1908) = 1.1 \pm 0.2$
implies very high densities, in the range 
$n(e) \simeq 5\,000$--25\,000\,cm$^{-3}$,
which however may not be unusual for 
starbursts at $z = 2 - 3$ (Brinchmann et al. 2008;
Liu et al. 2008; Hainline et al. 2009).

We searched for nebular O\,{\sc iii}]\,$\lambda\lambda 1660.809, 1666.150$
emission, but these lines are below the detection limit of our data. 
This is not surprising,
 given that the O\,{\sc iii}] doublet
is weaker than C\,{\sc iii}] by a factor of $\sim 7$  
in the LBG composite spectrum of Shapley et al. (2003),
and the C\,{\sc iii}] lines in the Cosmic Horseshoe 
are only detected at the $\sim 7 \sigma$ significance level
(Table~\ref{tab:CIII}).
Any He\,{\sc ii}\,$\lambda 1640.418$ emission
from Wolf-Rayet stars is also too weak to be identified
with certainty in our spectrum.


\section{Discussion}

We now consider the results of the previous sections 
in the light of some of the questions raised in the 
Introduction.

\subsection{Metallicity estimates}
With available instrumentation, only strongly lensed galaxies 
offer the means to cross-check at high redshift 
the metallicity of early-type stars with that of the 
H\,{\sc ii} regions that surround them.
We naturally expect the two measures to be the same,
since the stars presumably formed very recently
out of the gas which they now ionize; thus, this is
essentially a consistency check between different
metallicity indicators, all the more important given the
known systematic offsets between different emission line measures
of chemical abundance in H\,{\sc ii} regions (see, for
example, Kewley \& Ellison 2008). Such offsets in turn 
impact on the interpretation of chemical patterns and of
trends, such as the mass-metallicity relation 
and its evolution with redshift, which can give us insights
into galactic chemical evolution.

In Table~\ref{tab:Zs} we have collected available determinations
of element abundances in the Cosmic Horseshoe, bringing
together the stellar data from Section~\ref{sec:stellar_spectrum}
with the nebular measurements by Hainline et al. (2009).
The first three entries in the Table give the oxygen abundance
deduced from strong emission line indices, respectively
the $R23$ index of Pagel et al. (1979), for which
Hainline et al. (2009) used the calibration by Tremonti 
et al. (2004) based on observations of tens of thousands of
nearby galaxies in the \textit{SDSS}, and
the $N2$ and $O3N2$ indices of Pettini \& Pagel (2004).
The last three entries in the Table are from our 
analysis in this paper of UV spectral features from 
OB stars in the Horseshoe.

It is difficult to quote an error for each of the 
entries in Table~\ref{tab:Zs}; a realistic estimate 
should take into account the S/N of the data,
the scatter in the adopted index calibration and the
systematic bias which may affect it. However, most
estimates evidently converge towards a metallicity
of about a half solar (which, incidentally, is quite
typical of star-forming galaxies at these redshifts---see
Erb et al. 2006a). 
It is reassuring that stellar and nebular abundances
are generally in good mutual agreement. The only
deviant measure appears to be from the $R23$ method
which is known, however, to overestimate the oxygen abundance
in the near-solar regime, particularly
when the calibration
by Tremonti et al. (2004) is used 
(e.g. Kennicutt, Bresolin \& Garnett 2003;
Kewley \& Ellison 2008).

Less satisfactory aspects of the comparison in Table~\ref{tab:Zs}
are: (i) the poorly understood failure of the 1978 index
of Rix et al. (2004) 
to provide a meaningful estimate of metallicity 
in this particular case;
and (ii) the current limitations in the use of the 
wind lines as accurate abundance diagnostics.
Point (i) can only be addressed with a larger set 
of high quality UV spectra of high redshift galaxies.
Until it is resolved, however, it may be unwise to 
use this index to investigate the possibility
of a differential enrichment
of Fe-peak elements compared to the products of
Type-II supernovae (Pettini et al. 2002; Halliday et al. 2008). 
Concerning point (ii), it is frustrating that while the
strong wind lines are in principle one of the most 
easily measured abundance diagnostics---at least in
data of sufficient resolution  to resolve stellar and
interstellar components---we still lack a comprehensive
library of empirical ultraviolet spectra of OB stars 
of different metallicities to realise their full 
potential. The \textit{HST} survey of the Magellanic Clouds 
by Leitherer et al. (2001) went some way towards 
remedying  this situation, but their sampling of
the upper H-R diagram is still too sparse to 
assemble separate sets of Large and Small
Magellanic Cloud stars. The resulting hybrid
library, obtained by combining all the available
spectra into one set,
can only give an approximate measure of 
metallicity, given the factor of 
$\sim 2$ difference in the 
oxygen abundance of the two Clouds.

\begin{table}
 \begin{minipage}{152mm}
 \caption{\textsc{Metallicity Comparison}}
 \begin{tabular}{llll} 
\hline 
\hline
Method & Element(s) & $Z/Z_{\odot}^{\rm a}$ & Comments \\ 
\hline 
$R23$    & O                & 1.5                                   & H\,{\sc ii} regions$^{\rm b}$ \\
$N2$     & O                & 0.5                                  & H\,{\sc ii} regions$^{\rm b}$ \\
$O3N2$ & O               & 0.5                                   & H\,{\sc ii} regions$^{\rm b}$ \\
1425       & C, Si, Fe     & 0.5                                   & Photospheric, OB stars$^{\rm c}$\\
1978       & Fe              & \ldots                             & Photospheric, B stars$^{\rm c}$\\
C\,{\sc iv} & C, N, O, Fe & $\sim 0.6$                  & Stellar wind, O stars$^{\rm d}$\\
\hline
     \label{tab:Zs}
 \end{tabular}

$^{\rm a}$ Abundance relative to solar (on a linear scale),
using the \\
compilation of solar abundances by Asplund et al. (2005).\\
$^{\rm b}$ As reported by Hainline et al. (2009).\\
$^{\rm c}$ This work (Section~\ref{section:R04Indices}).\\
$^{\rm d}$ This work (Section~\ref{section:stellarwind}).\\
 \end{minipage}
 \end{table}

\subsection{Escape of ionizing photons}

It is still unclear what determines $f_{\rm esc}^{\rm LyC}$,
the fraction of hydrogen ionizing photons which escape
from star-forming galaxies into the intergalactic medium.
Direct detection of these Lyman continuum photons
has proved problematic until recently
(e.g. Shapley et al. 2006; Iwata et al. 2009),
and yet $f_{\rm esc}^{\rm LyC}$ must have
been large at very high redshifts for
the Universe to be reionized by the star-formation activity
thought to have taken place at $z > 6$
(e.g. Bolton \& Haehnelt 2007; Ryan-Weber et al. 2009).

Observations of gravitationally lensed galaxies may
offer insights into the factors that control 
$f_{\rm esc}^{\rm LyC}$. By fully resolving the 
interstellar absorption lines in our ESI spectrum of
the Cosmic Horseshoe, we reached the conclusion
that the interstellar gas only covers $\sim 60$\%
of the stellar UV light, as viewed from Earth.
This is a promising prerequisite for large
values of $f_{\rm esc}^{\rm LyC}$,
making the Horseshoe a high priority
candidate in searches for LyC emission 
from high-$z$ galaxies.
On the other hand, 
if $\sim 40$\% of the photons from 
the stars and surrounding H\,{\sc ii}
regions really had a clear path out of the galaxy,
we may have expected to see a strong and narrow \lya\ emission
line centred at $z_{\rm H\,\textsc{ii}}$, whereas no such 
feature is present in our spectrum.
Possibly, some 40\% of the stars 
are located behind neutral gas of too low a 
column density to
give discernible absorption in the metal lines,
and yet capable of scattering most of the 
\lya\  photons out of the line of sight 
(Hansen \& Oh 2006).
Whether such gas would be optically thick
to LyC photons remains to be established.

Another necessary condition for the escape of 
LyC photons is a weaker H$\alpha$ emission line than 
expected on the basis of the UV continuum luminosity
and reddening (admittedly in the idealised
case of continuous star formation at a constant rate).
Such a disparity would arise from a `matter-bounded nebula',
where not all of the LyC photons emitted by the stars
are absorbed and reprocessed within the H\,{\sc ii} region. 
As discussed in Section~\ref{sec:f_esc},
in the Cosmic Horseshoe SFR(H$\alpha) \nless$\,SFR(UV),
although a number of different factors, apart from
leakage of LyC photons, can affect this comparison.  
In conclusion, it would definitely be worthwhile
to search for LyC emission in the Cosmic Horseshoe
at wavelengths below 3085\,\AA\ (the 
redshifted value of the Lyman limit at $z = 2.38115$).
Furthermore, it would be of interest
to check for partial covering of the stars by the 
foreground interstellar medium in more galaxies
among the newly discovered strongly lensed sources,
ideally including galaxies with a range of \lya\ 
equivalent widths.
With such data in hand, we should be in a better position
to understand the conditions that determine the escape fraction
of LyC photons and the morphology of the \lya\ line.

\subsection{Comparison with MS\,1512-cB58}

One of the motivations for the present study was to 
establish how typical are the properties of the galaxy 
MS\,1512-cB58, the only previous case where the 
gravitational lensing boost was sufficient to allow
a detailed look at the spectrum of 
a high-redshift star-forming galaxy.
While we have now doubled the `sample'
of high-$z$ galaxies with good-quality ESI spectra,
it would clearly be premature to draw general
conclusions on the basis of just two objects.
Nevertheless, one cannot help but being struck by how closely
these two galaxies resemble each other in many of
their properties.
They are very similar in their overall metallicity and probably
in their detailed chemical composition,
indicating that they have reached comparable stages in
the conversion of their gas reservoirs into stars. Their young 
stellar populations are largely indistinguishable in their
UV spectral features, especially the P-Cygni lines which
are sensitive indicators of the mode of massive star formation
and of differential reddening among stars at the upper end of the IMF.
Most remarkable perhaps is the unexpected similarity in the 
properties of their outflowing interstellar media, 
with velocities spanning in both cases $\sim 1000$\,km~s$^{-1}$,
from $-800$ to $+250$\,km~s$^{-1}$ relative to the stars,
and involving comparable column densities of gas.

Presumably, it is just a coincidence that the first two galaxies
at $z = 2$--3 whose UV spectra have been
put under the `microscope' of gravitational
lensing have also turned out to be similar in so many respects.
After all, we now know that the Cosmic Horseshoe and 
MS\,1512-cB58 are also forming stars
at comparable rates of $\sim 50$--$100\,M_{\odot}$~yr$^{-1}$,
and have similar dynamical masses of 
$M_{\rm vir} \sim 1 \times 10^{10} M_\odot$
(Hainline et al. 2009 for these measurements
in  the Cosmic Horseshoe and Teplitz et al. 2000 for cB58).
Future studies of other gravitationally lensed objects will 
arguably show a wider range of properties of star-forming
galaxies at these redshifts.

What is interesting then, given the underlying similarity
of the Horseshoe and cB58, is to consider the aspects in
which they \emph{differ}: the incomplete coverage
of the stellar light by the interstellar absorption lines and the 
strikingly different \lya\ morphologies.
It is these two factors alone that would lead an observer
to `classify' the two galaxies differently. 
For example, in the rough breakdown of LBGs by Shapley et al. (2003)
according to their UV spectral properties, 
cB58 would fall in the quartile with the highest interstellar
absorption line equivalent widths and reddening,
while the Cosmic Horseshoe would be in the quartile with 
the second highest \lya\ \emph{emission} equivalent width.
And yet, from what we have uncovered in the present study,
it may be the case that these differences are perhaps simply
due to orientation effects---viewing two intrinsically similar
galaxies along differing sightlines through their interstellar
media, one (the Horseshoe) more rarefied and highly ionized
than the other (cB58). 
With future studies of other strongly lensed
galaxies we shall hopefully learn how to distinguish
intrinsic differences from those caused by
second-order effects, of which the direction
from which we view the central starburst may be one example.
For the moment, however, we must remain cautious of overinterpreting
differences between different `classes' of high-redshift galaxies,
especially if such distinctions are based on the morphology and
strength of the \lya\ line, whose appearance is so sensitive
to a variety of different parameters
(e.g. Pentericci et al. 2009).

\section{Summary and Conclusions}

We have presented ESI observations of the rest-frame
ultraviolet spectrum of a 
star-forming galaxy at $z = 2.38115$, lensed 
by a massive foreground galaxy into a nearly complete
Einstein ring dubbed the
`Cosmic Horseshoe'.
The high gravitational magnification  
affords the rare opportunity of recording
the spectrum of this 
member of the $z = 2$--3 galaxy population
at higher resolution and S/N than otherwise
achievable, and thereby provides us with a close-up look
of a galaxy during the epoch when cosmic star-formation
activity was at its peak.
Our main findings are as follows.

(i) The Cosmic Horseshoe shares many of its 
properties with the population of `BX' galaxies
selected with the colour criteria of Steidel et al. (2004).
Its metallicity $Z \simeq 0.5 Z_{\odot}$,
dynamical mass $M_{\rm vir} \simeq 1 \times 10^{10} M_\odot$,
and reddening $E(B-V) = 0.15$ are all typical of those
galaxies (Erb et al. 2006a,b,c).
With a star-formation rate SFR\,$\simeq 100 M_\odot$~yr$^{-1}$,
the Horseshoe is among the most luminous galaxies in the BX sample.

(ii) Generally, there is good agreement between different 
metallicity indicators based, respectively, on ratios of strong 
nebular emission lines, on blends of stellar photospheric lines,
and on P-Cygni lines from the most luminous OB stars.
The $R23$ index of Pagel et al. (1979) seems to overpredict
the oxygen abundance by a factor of $\sim 3$, when calibrated
with the prescription by Tremonti et al. (2004). 
Of the stellar measures,
the 1978 index of Rix et al. (2004)
does not fit the data well due to excess absorption
not present in the models and whose identity remains unclear.

(iii) A continuous mode of star formation
with a Salpeter slope for stars with masses 
$100 \geq M \geq 5\,M_\odot$
gives a good representation of the
UV spectrum; we see no evidence for a departure from 
a Salpeter IMF, nor differential reddening
among massive stars, in the contrast of the P-Cygni
lines over the underlying UV continuum.

(iv) The interstellar absorption lines are broad,
extending over a velocity interval 
of $\sim 1000$\,km~s$^{-1}$, from
$-800$ to $+250$\,km~s$^{-1}$ relative to
the redshift of the stars and H\,{\sc ii} regions.
The gas with the highest optical depths is outflowing
from the regions of star formation with a speed 
of $\sim 150$\,km~s$^{-1}$, 
and only covers $\sim 60$\%
of the stellar continuum light.

(v) The \lya\ line is in emission
and shares many of the properties
of the so-called \lya\ emitters,
although its equivalent width is a
factor of $\sim 2$ smaller than the 
lower limit
$W_0({\rm Ly}\alpha) \geq 20$\,\AA\
commonly adopted to define this subset 
of galaxies.
The resolved line profile 
matches well those computed with models
of resonantly scattered \lya\ photons in an expanding
medium, although such models find it difficult
to account for the fact that the velocity dispersion of
the gas is comparable to its bulk expansion velocity, and
for the existence of gas extending over hundreds of km~s$^{-1}$
in velocity space. 
Some $\sim 10$--15\% of the \lya\ photons produced
escape the galaxy, the remainder presumably 
being absorbed by dust
and recycled into infrared photons,
or scattered over a larger area than that
covered by our narrow-slit observations.
It remains to be established what 
fraction of the Lyman \textit{continuum}
photons emitted by the stars
leak into the IGM.

(vi) Overall, many of the physical properties of the stars and
interstellar medium of the Cosmic Horseshoe are similar to those
of the only other galaxy previously studied at comparable resolution
and S/N, MS\,1512-cB58 at $z = 2.7276$. This may not be surprising,
given that both galaxies are fairly typical examples of the star-forming
galaxy population at these redshifts.
The fundamental similarities between the two objects
highlight the fact that the aspects in which they differ,
particularly the very different morphologies of their \lya\
lines, may be due to superficial reasons,
such as differing viewing angles, rather than to
more deeply rooted causes.

In the near future it should be possible to make further progress
on all of these topics with the increasing numbers of strongly 
lensed high-redshift galaxies being discovered by the \textit{SDSS}
and other lensing surveys.

\section*{Acknowledgements}
We are grateful to the staff at the W.~M. Keck Observatory for their
competent assistance with the observations, to Sam Rix and Dan Nestor who 
generously provided some of the software used in the analysis of the spectra,
and to Simon Dye for clarifications concerning the lensing models.
Lindsay King kindly allowed us to reproduce her ESO/FORS2 image
of the Cosmic Horseshoe in Figure~\ref{fig:horseimage}.
We also thank the anonymous referee for helpful comments 
which have improved the paper.
AMQ's research is funded by a scholarship from the Marshall Foundation.
AES acknowledges support from the
David and Lucile Packard Foundation and the Alfred P. Sloan
Foundation, and CCS from NSF grant AST-0606912 and the
John D. and Catherine T. MacArthur Foundation.
Finally, we wish to extend thanks to those of 
Hawaiian ancestry on whose mountain
we are privileged to be guests.

\label{lastpage}


\begin{thebibliography}{}
\small
\itemindent -0.48cm

\bibitem[]{}Asplund, M., Grevesse, N, \& Sauval, A.~J.\ 2005,
in Barnes T.~G.~III, \& Bash, F.~N. eds., 
ASP Conf. Ser. Vol. 336, Cosmic Abundances as Records
of Stellar Evolution and Nucleosynthesis.
Astron. Soc. Pac., San Francisco, p.~25

\bibitem[Baker et al.(2004)]{2004ApJ...604..125B} Baker, A.~J., Tacconi, 
L.~J., Genzel, R., Lehnert, M.~D., \& Lutz, D.\ 2004, \apj, 604, 125 

\bibitem[Baker et al.(2001)]{2001A&A...372L..37B} 
Baker, A.~J., Lutz, D., Genzel, R., Tacconi, L.~J., \& Lehnert, M.~D.\ 2001, \aap, 372, L37 

\bibitem[Belokurov et al.(2007)]{2007ApJ...671L...9B} Belokurov, V., et 
al.\ 2007, \apjl, 671, L9 

\bibitem[Belokurov et al.(2009)]{2009MNRAS.392..104B} 
Belokurov, V., Evans,  N.~W., Hewett, P.~C., Moiseev, A., 
McMahon, R.~G., Sanchez, S.~F., \& King, L.~J.\ 2009, \mnras, 392, 104 

\bibitem[Bolton \& Haehnelt(2007)]{2007MNRAS.382..325B} 
Bolton, J.~S., \& Haehnelt, M.~G.\ 2007, \mnras, 382, 325 

\bibitem[Brinchmann et al.(2008)]{2008MNRAS.385..769B} 
Brinchmann, J., Pettini, M., \& Charlot, S.\ 2008, \mnras, 385, 769 

\bibitem[Calzetti et al.(2000)]{2000ApJ...533..682C} 
Calzetti, D., Armus, 
L., Bohlin, R.~C., Kinney, A.~L., Koornneef, J., 
\& Storchi-Bergmann, T.\ 2000, \apj, 533, 682 

\bibitem[Chabrier(2003)]{2003PASP..115..763C} 
Chabrier, G.\ 2003, \pasp, 115, 763 

\bibitem[Chandar et al.(2005)]{2005ApJ...628..210C} Chandar, R., Leitherer, 
C., Tremonti, C.~A., Calzetti, D., Aloisi, A., Meurer, G.~R., 
\& de Mello, D.\ 2005, \apj, 628, 210 

\bibitem[Crowther et al.(2006)]{2006MNRAS.368..895C} Crowther, P.~A., 
Prinja, R.~K., Pettini, M., \& Steidel, C.~C.\ 2006, \mnras, 368, 895 

\bibitem[Dye et al.(2008)]{2008MNRAS.388..384D} 
Dye, S., Evans, N.~W., Belokurov, V., Warren, S.~J., 
\& Hewett, P.\ 2008, \mnras, 388, 384 

\bibitem[Ellingson et al.(1996)]{1996ApJ...466L..71E} Ellingson, E., Yee, 
H.~K.~C., Bechtold, J., \& Elston, R.\ 1996, \apjl, 466, L71 

\bibitem[Erb et al.(2006a)]{2006ApJ...644..813E} Erb, D.~K., Shapley, A.~E., 
Pettini, M., Steidel, C.~C., Reddy, N.~A., 
\& Adelberger, K.~L.\ 2006a, \apj, 644, 813 

\bibitem[Erb et al.(2006b)]{2006ApJ...646..107E} 
Erb, D.~K., Steidel, C.~C., Shapley, A.~E., Pettini, M., Reddy, N.~A., 
\& Adelberger, K.~L.\ 2006b, \apj, 646, 107 

\bibitem[Erb et al.(2006c)]{2006ApJ...647..128E} 
Erb, D.~K., Steidel, C.~C., 
Shapley, A.~E., Pettini, M., Reddy, N.~A., 
\& Adelberger, K.~L.\ 2006c, \apj, 647, 128 

\bibitem[Faucher-Gigu{\`e}re et al.(2008)]{2008ApJ...682L...9F} 
Faucher-Gigu{\`e}re, C.-A., Lidz, A., Hernquist, L., 
\& Zaldarriaga, M.\ 2008, \apjl, 682, L9 

\bibitem[Finkelstein et al.(2009)]{2009arXiv0905.1122F} Finkelstein, S.~L., 
Papovich, C., Rudnick, G., Egami, E., Le Floc'h, E., Rieke, M.~J., Rigby, 
J., \& Willmer, C.~N.~A.\ 2009,  \apj, in press

\bibitem[Grimes et al.(2009)]{2009ApJS..181..272G} 
Grimes, J.~P., et al.\ 2009, \apjs, 181, 272 

\bibitem[]{} Hainline, K.~N., Shapley, A.~E., Kornei, K.~A.,
Pettini, M., Buckley-Geer, E., Allam, S.~S., \& Tucker, D.~L. \ 
2009, \apj, submitted

\bibitem[Halliday et al.(2008)]{2008A&A...479..417H} 
Halliday, C., et al.\ 2008, \aap, 479, 417 

\bibitem[Hansen \& Oh(2006)]{2006MNRAS.367..979H} 
Hansen, M., \& Oh, S.~P.\ 2006, \mnras, 367, 979 

\bibitem[Heckman et al.(2000)]{2000ApJS..129..493H} Heckman, T.~M., 
Lehnert, M.~D., Strickland, D.~K., \& Armus, L.\ 2000, \apjs, 129, 493 

\bibitem[Hu et al.(1998)]{1998ApJ...502L..99H} Hu, E.~M., Cowie, L.~L., 
\& McMahon, R.~G.\ 1998, \apjl, 502, L99

\bibitem[Iwata et al.(2009)]{2009ApJ...692.1287I} 
Iwata, I., et al.\ 2009, \apj, 692, 1287 

\bibitem[Jenkins \& Tripp(2006)]{2006ApJ...637..548J} 
Jenkins, E.~B., \& Tripp, T.~M.\ 2006, \apj, 637, 548 

\bibitem[Kennicutt(1998)]{1998ARA&A..36..189K} 
Kennicutt, R.~C., Jr.\ 1998, \araa, 36, 189 

\bibitem[Kennicutt et al.(2003)]{2003ApJ...591..801K} 
Kennicutt, R.~C., Jr., Bresolin, F., \& Garnett, D.~R.\ 2003, \apj, 591, 801 

\bibitem[Kewley \& Ellison(2008)]{2008ApJ...681.1183K} 
Kewley, L.~J., \& Ellison, S.~L.\ 2008, \apj, 681, 1183 

\bibitem[Kobulnicky \& Kewley(2004)]{2004ApJ...617..240K} 
Kobulnicky, H.~A., \& Kewley, L.~J.\ 2004, \apj, 617, 240 

\bibitem[Kubo et al.(2009)]{2008arXiv0812.3934K} 
Kubo, J.~M., Allam, S.~S., 
Annis, J., Buckley-Geer, E.~J., Diehl, H.~T., Kubik, D., Lin, H., 
\& Tucker, D.\ 2009, arXiv:0812.3934 

\bibitem[Kudritzki \& Puls(2000)]{2000ARA&A..38..613K} 
Kudritzki, R.-P., \& Puls, J.\ 2000, \araa, 38, 613 

\bibitem[]{} Leitherer, C. \ 2008, in Hunt, K.~L., Madden, S., 
\& Schneider, R., eds., IAU Symp. 255,
Low-Metallicity Star Formation: From the First Stars
to Dwarf Galaxies.
Cambridge Univ. Press, Cambridge, p.\,305

\bibitem[Leitherer et al.(1999)]{1999ApJS..123....3L} Leitherer, C., et 
al.\ 1999, \apjs, 123, 3 

\bibitem[Leitherer et al.(2002)]{2002ApJ...574..114L} Leitherer, C., 
Calzetti, D., \& Martins, L.~P.\ 2002, \apj, 574, 114 

\bibitem[Leitherer et al.(2001)]{2001ApJ...550..724L} Leitherer, C., 
Le{\~a}o, J.~R.~S., Heckman, T.~M., Lennon, D.~J., Pettini, M., 
\& Robert, C.\ 2001, \apj, 550, 724 

\bibitem[Liu et al.(2008)]{2008ApJ...678..758L} Liu, X., Shapley, A.~E., 
Coil, A.~L., Brinchmann, J., \& Ma, C.-P.\ 2008, \apj, 678, 758 

\bibitem[Mar \& Bailey(1995)]{1995PASA...12..239M} 
Mar, D.~P., \& Bailey, G.\ 1995, 
Publications of the Astronomical Society of Australia, 12, 239 

\bibitem[Martin(2005)]{2005ApJ...621..227M} 
Martin, C.~L.\ 2005, \apj, 621, 227 

\bibitem[]{} Martin, C.~L. \& Bouch{\'e}, N.\ 2009,
\apj, submitted

\bibitem[Mas-Hesse et al.(2003)]{2003ApJ...598..858M} 
Mas-Hesse, J.~M., Kunth, D., Tenorio-Tagle, G., Leitherer, C., 
Terlevich, R.~J., \& Terlevich, E.\ 2003, \apj, 598, 858 

\bibitem[Morton(2003)]{2003ApJS..149..205M} 
Morton, D.~C.\ 2003, \apjs, 149, 205 

\bibitem[Neufeld(1991)]{1991ApJ...370L..85N} 
Neufeld, D.~A.\ 1991, \apjl, 370, L85 

\bibitem[Pagel(2003)]{2003ASPC..304..187P} 
Pagel, B.~E.~J.\ 2003,  in Charbonnel, C., Schaerer, D., 
\& Meynet, G., eds., 
CNO Abundances in Dwarf and Spiral Galaxies,
Astronomical Society of the Pacific Conference Series, 
San Francisco, 304, p. 187 

\bibitem[Pagel et al.(1979)]{1979MNRAS.189...95P} 
Pagel, B.~E.~J., Edmunds, M.~G., Blackwell, D.~E., 
Chun, M.~S., \& Smith, G.\ 1979, \mnras, 189, 95 

\bibitem[Papovich et al.(2001)]{2001ApJ...559..620P} Papovich, C., 
Dickinson, M., \& Ferguson, H.~C.\ 2001, \apj, 559, 620 

\bibitem[Partridge \& Peebles(1967)]{1967ApJ...147..868P} 
Partridge, R.~B., \& Peebles, P.~J.~E.\ 1967, \apj, 147, 868 

\bibitem[]{} Pentericci, L., Grazian, A., Fontana, A., Castellano, M.,
Giallongo, E., Salinbeni, S., \& Santini, P. \ 2009, 
\aap, in press ( arXiv:0811.1861)

\bibitem[]{}Pettini, M. 2006 in
LeBrun V.,  Mazure A.,  Arnouts S. \& Burgarella D., eds.,
The Fabulous Destiny of Galaxies: Bridging Past and Present.
Frontier Group, Paris, p. 319 
(astro-ph/0603066).

\bibitem[Pettini et al.(2007)]{2007NCimB.122.1043P} 
Pettini, M., et al.\  2007, Nuovo Cimento B Serie, 122, 1043 

\bibitem[Pettini \& Pagel(2004)]{2004MNRAS.348L..59P} 
Pettini, M., \& Pagel, B.~E.~J.\ 2004, \mnras, 348, L59 

\bibitem[Pettini et al.(2002)]{2002ApJ...569..742P} Pettini, M., Rix, 
S.~A., Steidel, C.~C., Adelberger, K.~L., Hunt, M.~P., 
\& Shapley, A.~E.\ 2002, \apj, 569, 742 

\bibitem[Pettini et al.(2001)]{2001ApJ...554..981P} Pettini, M., Shapley, 
A.~E., Steidel, C.~C., Cuby, J.-G., Dickinson, M., Moorwood, A.~F.~M., 
Adelberger, K.~L., \& Giavalisco, M.\ 2001, \apj, 554, 981 

\bibitem[Pettini et al.(2000)]{2000ApJ...528...96P} Pettini, M., Steidel, 
C.~C., Adelberger, K.~L., Dickinson, M., 
\& Giavalisco, M.\ 2000, \apj, 528, 96 

\bibitem[Reddy \& Steidel(2009)]{2009ApJ...692..778R} 
Reddy, N.~A., \& Steidel, C.~C.\ 2009, \apj, 692, 778 

\bibitem[Reddy et al.(2008)]{2008ApJS..175...48R} Reddy, N.~A., Steidel, 
C.~C., Pettini, M., Adelberger, K.~L., Shapley, A.~E., Erb, D.~K., 
\& Dickinson, M.\ 2008, \apjs, 175, 48 

\bibitem[Rhoads et al.(2000)]{2000ApJ...545L..85R} Rhoads, J.~E., Malhotra, 
S., Dey, A., Stern, D., Spinrad, H., 
\& Jannuzi, B.~T.\ 2000, \apjl, 545, L85 

\bibitem[Rix et al.(2004)]{2004ApJ...615...98R} 
Rix, S.~A., Pettini, M., Leitherer, C., Bresolin, F., Kudritzki, R.-P., 
\& Steidel, C.~C.\ 2004, \apj, 615, 98 

\bibitem[Ryan-Weber et al.(2009)]{2009arXiv0902.1991R} 
Ryan-Weber, E.~V.,  Pettini, M., Madau, P., \& Zych, B.~J.\ 2009, 
\mnras, in press (arXiv:0902.1991)

\bibitem[Saito et al.(2006)]{2006ApJ...648...54S} 
Saito, T., Shimasaku, K.,  Okamura, S., Ouchi, M., 
Akiyama, M., \& Yoshida, M.\ 2006, \apj, 648, 54 

\bibitem[Salpeter(1955)]{1955ApJ...121..161S} 
Salpeter, E.~E.\ 1955, \apj, 121, 161 

\bibitem[]{} Savage, B.~D. \& Sembach, K.~R.\ 1991, \apj, 379, 245

\bibitem[Savaglio et al.(2002)]{2002ApJ...567..702S} 
Savaglio, S., Panagia, N., \& Padovani, P.\ 2002, \apj, 567, 702 

\bibitem[Sawicki(2001)]{2001AJ....121.2405S} 
Sawicki, M.\ 2001, \aj, 121, 2405 


\bibitem[Schaerer(2007)]{2007arXiv0706.0139S} 
Schaerer, D.\ 2007, in  Cepa, J. ed,
The Emission Line Universe, Cambridge Univ. Press, Cambridge,
in press (arXiv:0706.0139 )

\bibitem[Seitz et al.(1998)]{1998MNRAS.298..945S} Seitz, S., Saglia, R.~P., 
Bender, R., Hopp, U., Belloni, P., \& Ziegler, B.\ 1998, \mnras, 298, 945 

\bibitem[Shapley et al.(2001)]{2001ApJ...562...95S} Shapley, A.~E., 
Steidel, C.~C., Adelberger, K.~L., Dickinson, M., Giavalisco, M., 
\& Pettini, M.\ 2001, \apj, 562, 95 

\bibitem[Shapley et al.(2005)]{2005ApJ...626..698S} Shapley, A.~E., 
Steidel, C.~C., Erb, D.~K., Reddy, N.~A., Adelberger, K.~L., Pettini, M., 
Barmby, P., \& Huang, J.\ 2005, \apj, 626, 698 

\bibitem[Shapley et al.(2003)]{2003ApJ...588...65S} Shapley, A.~E., 
Steidel, C.~C., Pettini, M., \& Adelberger, K.~L.\ 2003, \apj, 588, 65 

\bibitem[Shapley et al.(2006)]{2006ApJ...651..688S} 
Shapley, A.~E., Steidel, C.~C., Pettini, M., Adelberger, K.~L., 
\& Erb, D.~K.\ 2006, \apj, 651, 688 

\bibitem[Sheinis et al.(2000)]{2000SPIE.4008..522S} Sheinis, A.~I., Miller, 
J.~S., Bolte, M., \& Sutin, B.~M.\ 2000, \procspie, 4008, 522 

\bibitem[Siana et al.(2008)]{2008ApJ...689...59S} Siana, B., Teplitz, 
H.~I., Chary, R.-R., Colbert, J., \& Frayer, D.~T.\ 2008, \apj, 689, 59 

\bibitem[Smail et al.(2007)]{2007ApJ...654L..33S} 
Smail, I., et al.\ 2007,  \apj, 654, L33 

\bibitem[Stark et al.(2008)]{2008Natur.455..775S} 
Stark, D.~P., Swinbank, A.~M., Ellis, R.~S., Dye, S., Smail, I.~R., 
\& Richard, J.\ 2008, \nat, 455, 775 


\bibitem[Steidel et al.(1999)]{1999ApJ...519....1S} 
Steidel, C.~C.,  Adelberger, K.~L., Giavalisco, M., Dickinson, M., 
\& Pettini, M.\ 1999, \apj, 519, 1

\bibitem[Steidel et al.(2004)]{2004ApJ...604..534S} Steidel, C.~C., 
Shapley, A.~E., Pettini, M., Adelberger, K.~L., Erb, D.~K., Reddy, N.~A., 
\& Hunt, M.~P.\ 2004, \apj, 604, 534 

 
\bibitem[Tapken et al.(2007)]{2007A&A...467...63T} 
Tapken, C., Appenzeller, I., Noll, S., Richling, S., 
Heidt, J., Meink{\"o}hn, E., \& Mehlert, D.\ 2007, \aap, 467, 63 

\bibitem[Teplitz et al.(2000)]{2000ApJ...533L..65T} 
Teplitz, H.~I., et al.\ 2000, \apjl, 533, L65 

\bibitem[Tremonti et al.(2004)]{2004ApJ...613..898T} 
Tremonti, C.~A., et al.\ 2004, \apj, 613, 898 

\bibitem[Vanzella et al.(2009)]{2009arXiv0901.4364V} Vanzella, E., et al.\ 
2009, \apj, in press (arXiv:0901.4364)

\bibitem[Verhamme et al.(2006)]{2006A&A...460..397V} 
Verhamme, A., Schaerer, D., \& Maselli, A.\ 2006, \aap, 460, 397 

\bibitem[Verhamme et al.(2008)]{2008A&A...491...89V} 
Verhamme, A., Schaerer, D., Atek, H., \& Tapken, C.\ 2008, \aap, 491, 89 

\bibitem[Wilkins et al.(2008)]{2008MNRAS.391..363W} Wilkins, S.~M., 
Hopkins, A.~M., Trentham, N., \& Tojeiro, R.\ 2008, \mnras, 391, 363 

\bibitem[Yee et al.(1996)]{1996AJ....111.1783Y} 
Yee, H.~K.~C., Ellingson, E., Bechtold, J., Carlberg, R.~G., 
\& Cuillandre, J.-C.\ 1996, \aj, 111, 1783 

\bibitem[Yuan \& Kewley(2009)]{2009arXiv0906.0371Y} 
Yuan, T.-T., \& Kewley, L.~J.\ 2009, ApJ, in press


\end{thebibliography}
\end{document}